\documentclass[10pt]{iopart}

\usepackage{graphicx} 
\usepackage{cite}     

\begin{document}

\newcommand{\dd}{{\rm{d}}} 
\newcommand{\rovno}{& = &} 
\newcommand{\ssqrt}{{\textstyle\frac1{\sqrt{2}}}}          
\newcommand{\boldu}{\mbox{\boldmath$u$}}                   
\newcommand{\bolde}{\mbox{\boldmath$e$}}                   
\newcommand{\boldk}{\mbox{\boldmath$k$}}                   
\newcommand{\boldl}{\mbox{\boldmath$l$}}                   
\newcommand{\boldm}{\mbox{\boldmath$m$}}                   
\newcommand{\bboldm}{\bar{\mbox{\boldmath$m$}}}            
\newcommand{\boldZ}{\mbox{\boldmath$Z$}}                   
\newcommand{\boldT}{\mbox{\boldmath$T$}}                   
\newcommand{\boldF}{\mbox{\boldmath$F$}}                   
\newcommand{\nat}{{\mathrm{nat}}}


\title[Physical interpretation of Kundt spacetimes using geodesic deviation]{Physical interpretation of Kundt spacetimes using geodesic deviation}

\author{J Podolsk\'y$^1$ and R \v{S}varc$^{1,2}$ }

\address{$^1$ Institute of Theoretical Physics, Faculty of Mathematics and Physics,
Charles University in Prague, V~Hole\v{s}ovi\v{c}k\'ach~2, 180~00 Praha 8, Czech Republic }
\address{$^2$ Department of Physics, Faculty of Science,
J.~E.~Purkinje University in {\'U}st{\'i} nad Labem, {\v C}esk{\'e} ml{\'a}de{\v z}e~8,
400~96 {\'U}st{\'i} nad Labem, Czech Republic}
\eads{\mailto{podolsky@mbox.troja.mff.cuni.cz} and \mailto{robert.svarc@mff.cuni.cz}}

\begin{abstract}
We investigate the fully general class of non-expanding, non-twisting and shear-free $D$-dimensional geometries using the invariant form of geodesic deviation equation which describes the relative motion of free test particles. We show that the local effect of such gravitational fields on the particles basically consists of isotropic motion caused by the cosmological constant $\Lambda$, Newtonian-type tidal deformations typical for spacetimes of algebraic type~D or~II, longitudinal motion characteristic for spacetimes of type~III, and type~N purely transverse effects of exact gravitational waves with ${D(D-3)/2}$ polarizations. We explicitly discuss the canonical forms of the geodesic deviation motion in all algebraically special subtypes of the Kundt family for which the optically privileged direction is a multiple Weyl aligned null direction (WAND), namely D(a), D(b), D(c), D(d), III(a), III(b), III$_i$, II$_i$, II(a), II(b), II(c) and II(d). We demonstrate that the key invariant quantities determining these algebraic types and subtypes also directly determine the specific local motion of test particles, and are thus measurable by gravitational detectors. As an example, we analyze an interesting class of type~N or~II gravitational waves which propagate on backgrounds of type~O or~D, including Minkowski, Bertotti--Robinson, Nariai and Pleba\'{n}ski--Hacyan universes.
\end{abstract}

\submitto{\CQG}


\maketitle

\section{Introduction}
\label{intro}

Spacetimes of the Kundt class are defined by a purely geometric property, namely that they admit a geodesic null congruence which is non-expanding, non-twisting and shear-free. In the context of four-dimensional general relativity, such vacuum and pure radiation spacetimes of type~N, III, or~O were introduced and initially studied 50~years ago by Wolfgang Kundt \cite{Kundt:1961,Kundt:1962}.

The whole Kundt class is, in fact, much wider. It admits a cosmological constant, electromagnetic field, other matter fields and supersymmetry. The solutions may be of various algebraic types and can be extended to any number~$D$ of dimensions. All Kundt spacetimes (without assuming field equations) can be written as
\begin{equation}
\hspace{-15mm}
\dd s^2 = g_{pq}(u,x)\, \dd x^p\dd x^q+2\,g_{up}(r,u,x)\, \dd u\,\dd x^p-2\,\dd u\,\dd r+g_{uu}(r,u,x)\,\dd u^2 \, , \label{obecny Kundtuv prostorocas}
\end{equation}
see \cite{Kundt:1961,Kundt:1962,Stephani:2003,GriffithsPodolsky:2009,PodolskyZofka:2009,ColeyEtal:2009,Coley:2008,OrtaggioPravdaPravdova:2013,PodolskySvarc:2013}.
In this metric, the coordinate $r$ is an affine parameter along the optically privileged null congruence ${\mathbf{k}=\partial_r}$ (with vanishing expansion, twist and shear), ${u=\ }$const. label null (wave)surfaces, and ${x\equiv(x^2, x^3, \ldots, x^{D-1})}$ are ${D-2}$ spatial coordinates in the transverse Riemannian space. The spatial part $g_{pq}$ of the metric must be independent of $r$, all other metric components $g_{up}$ and $g_{uu}$ can be functions of all the coordinates $(r,u,x)$.

The Kundt class of spacetimes is one of the most important families of exact solutions in Einstein's general relativity theory, see chapter~31 of the monograph \cite{Stephani:2003} or chapter~18 of \cite{GriffithsPodolsky:2009} for reviews of the standard ${D=4}$ case. It contains several famous subclasses, both in four and higher number of dimensions, with interesting mathematical and physical properties. The best-known of these are pp-waves (see \cite{Stephani:2003,GriffithsPodolsky:2009,Coley:2008,OrtaggioPravdaPravdova:2013,Bri25,ColMilPelPraPraZal03,ColeyMilsonPravdaPravdova:2004,ColMilPraPra04,ColFusHerPel06} and references therein) which admit a covariantly constant null vector field. There are also VSI and CSI spacetimes \cite{ColMilPelPraPraZal03,ColeyMilsonPravdaPravdova:2004,ColMilPraPra04,ColHerPel06,ColFusHerPel06,Coley:2008,ColHerPel09,OrtaggioPravdaPravdova:2013} for which all polynomial scalar invariants constructed from the Riemann tensor and its derivatives vanish and are constant, respectively. Moreover, all the relativistic gyratons known so far
\cite{Bon70,FroFur05,FroIsrZel05,FroZel05,FroZel06,CalLeZor07,KadlecovaZelnikovKrtousPodolsky:2009,KrtousPodolskyZelnikovKadlecova:2012}, representing the fields of localised spinning sources that propagate with the speed of light, are also specific members of the Kundt class. Vacuum and conformally flat pure radiation Kundt spacetimes provide an exceptional case for the invariant classification of exact solutions \cite{NuttMilsonColey:2013,Wils:1989,KoutrasMcIntosh:1996,EdgarLudwig:1997a,Skea:1997,GriffithsPodolsky:1998,Barnes:2001}, and all type~D pure radiation solutions are also known \cite{WilsVandenBergh:1990,GrooteBerghWylleman:2010}. All vacuum Kundt solutions of type~D were found and classified a long time ago \cite{Kinnersley:1969a} and generalized to electrovacuum and any value of the cosmological constant \cite{PlebanskiDemianski:1976,GriffithsPodolsky:2006b}. These contain a subfamily of direct-product spacetimes, namely the Bertotti--Robinson, (anti-)Nariai and Pleba\'{n}ski--Hacyan spacetimes of type O and D (see chapter~7 of \cite{GriffithsPodolsky:2009}, \cite{KadlecovaZelnikovKrtousPodolsky:2009} and \cite{KrtousPodolskyZelnikovKadlecova:2012} for higher-dimensional generalizations) representing, e.g., extremal limits and near-horizon geometries. With Minkowski and (anti-)de~Sitter spaces they form the natural backgrounds for non-expanding gravitational waves of types~N and~II \cite{OzsvathRobinsonRozga:1985,Siklos:1985,Podolsky:1998a,BicakPodolsky:1999a,BicakPodolsky:1999b,GriffithsDochertyPodolsky:2004,PodolskyOrtaggio:2003,PodolskyBelan:2004}.

In our studies here we consider the fully general class of Kundt spacetimes of an arbitrary  dimension ${D\ge4}$ (results for the standard general relativity are obtained by simply setting ${D=4}$). Taking the spacetime dimension as a free parameter $D$, we can investigate whether the extension of the Kundt family to ${D>4}$ exhibits some qualitatively different features and unexpected properties. Our paper is thus also a contribution to the contemporary research analyzing various aspects of Einstein's gravity extended to higher dimensions. Explicit Kundt solutions help us illustrate specific physical properties and general mathematical features of such theories.

Specifically, we systematically investigate the complete ${D\ge4}$ Kundt class of solutions using geodesic deviation and discuss the corresponding effects on free test particles. In section~\ref{geodevKundt} we summarize the equation of geodesic deviation, introduce invariant amplitudes of the gravitational field, and we discuss them for the fully general Kundt family of geometries. In section~\ref{specfrmes} we derive expressions for these amplitudes, and in  section~\ref{claasifKundt} we evaluate them explicitly for all algebraically special Kundt spacetimes for which the optically privileged congruence is generated by a multiple WAND. The main results are presented in section~\ref{subclasses} where we discuss the specific structure of relative motion of test particles for all possible algebraic types and subtypes of such Kundt geometries, see subsections~\ref{typO}--\ref{typII}. In the final section~\ref{example} we present a particular example, namely an interesting class of type~II and~N non-expanding gravitational waves on~D and~O backgrounds of any dimension.

\section{Geodesic deviation in the fully general Kundt spacetime}
\label{geodevKundt}

\begin{figure}
  \begin{center}
   \vspace{-1mm}
  \includegraphics[width=0.61\textwidth]{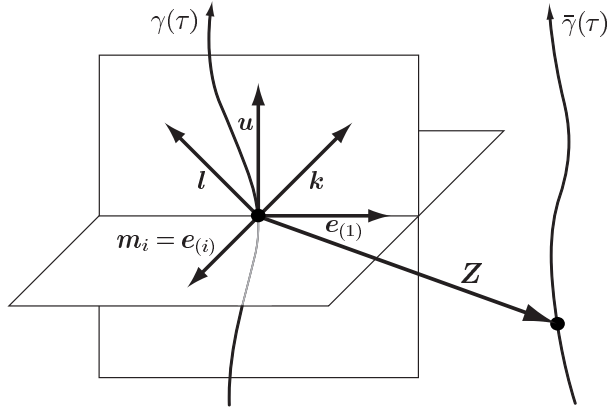}
  \vspace{-6mm}
  \end{center}
  \caption{\label{figure1} Evolution of the separation vector $\boldZ$ that connects particles moving along  geodesics $\gamma(\tau)$, $\bar\gamma(\tau)$ is given by the equation of geodesic deviation (\ref{invariant form of eq of geodesic deviationL}) and (\ref{invariant form of eq of geodesic deviationT}). Its components are expressed in the orthonormal frame ${\{\bolde_a \}}$ with ${\bolde_{(0)}=\boldu}$. The associated null frame ${\{\boldk, \boldl, \boldm_{i} \}}$ is also indicated.}
\end{figure}

Relative motion of nearby free test particles (without charge and spin) is described by the equation of geodesic deviation \cite{LeviCivita:1926,Synge:1934}. It has long been used as an important tool for studies of four-dimensional general relativity, in particular to analyze fields representing gravitational waves and black hole spacetimes (see \cite{Pirani:1956,Szekeres:1965,FeliceBini:book,PodolskySvarc:2012} for more details and references). In our recent work \cite{PodolskySvarc:2012} generalizing \cite{BicakPodolsky:1999b} we demonstrated that the equation of geodesic deviation in \emph{any} $D$-dimensional spacetime can be expressed in the invariant form (using Einstein's field equations ${R_{ab}-\frac{1}{2}R\, g_{ab} + \Lambda\,g_{ab}=8\pi\,T_{ab}}$)
\begin{eqnarray}
&&\hspace{-14mm}\ddot{Z}^{(1)} = \frac{2\Lambda}{(D-2)(D-1)}\,Z^{(1)} +\Psi_{2S}\,Z^{(1)}
+ \frac{1}{\sqrt{2}}\,(\,\Psi_{1T^j}-\Psi_{3T^j})\,Z^{(j)} \nonumber \\
 && \hspace{-2.5mm} + \, \frac{8\pi}{D-2}\!\left[T_{(1)(1)}\,Z^{(1)}+T_{(1)(j)}\,Z^{(j)}-\Big(T_{(0)(0)}+\frac{2\,T}{D-1}\Big)Z^{(1)}\right]\!, \label{invariant form of eq of geodesic deviationL}\\
&&\hspace{-13.8mm}\ddot{Z}^{(i)} = \frac{2\Lambda}{(D-2)(D-1)}\,Z^{(i)}-\Psi_{2T^{(ij)}}\,Z^{(j)}
+ \frac{1}{\sqrt{2}}\,(\,\Psi_{1T^i}-\Psi_{3T^i})\,Z^{(1)}\nonumber \\
 && \hspace{+55.8mm} -\frac{1}{2}\,(\,\Psi_{0^{ij}}+\Psi_{4^{ij}})\,Z^{(j)}\nonumber\\
 && \hspace{-2.5mm} + \, \frac{8\pi}{D-2}\!\left[T_{(i)(1)}\,Z^{(1)}+T_{(i)(j)}\,Z^{(j)}-\Big(T_{(0)(0)}+\frac{2\,T}{D-1}\Big)Z^{(i)}\right]\!, \label{invariant form of eq of geodesic deviationT}
\end{eqnarray}
${\,i,j=2,\ldots,D-1\,}$. Here ${Z^{(1)}, Z^{(2)}, \ldots, Z^{(D-1)}}$ are spatial components
${Z^{(\rm{i})} \equiv \bolde^{(\rm{i})}\cdot \boldZ}$
of the separation vector ${\boldZ}$ between the two test particles in a natural interpretation orthonormal frame ${\{\bolde_a \}}$, ${\bolde_a \cdot \bolde_b=\eta_{ab}}$, where ${\bolde_{(0)}=\boldu}$ is the velocity vector of the fiducial test particle, and ${\ddot{Z}^{(1)}, \ddot{Z}^{(2)}, \ldots, \ddot{Z}^{(D-1)}}$ are the corresponding relative physical accelerations
${\ddot Z^{(\rm{i})} \equiv \bolde^{(\rm{i})}\cdot ({{\rm D}^2\boldZ}/{\dd \tau^2})}$. The coefficients
${T_{ab}\equiv \boldT(\bolde_a,\bolde_b)}$ denote frame components of the energy-momentum tensor ($T$ is its trace), and the scalars
\begin{eqnarray}
\Psi_{0^{ij}} \rovno C_{abcd}\; k^a\, m_i^b\, k^c\, m_j^d \, , \nonumber \\
\Psi_{1T^{i}} \rovno C_{abcd}\; k^a\, l^b\, k^c\, m_i^d \, ,
\hspace{10mm} \Psi_{1^{ijk}} = C_{abcd}\; k^a\, m_i^b\, m_j^c\, m_k^d    \ ,\nonumber \\
\Psi_{2S} \rovno C_{abcd}\; k^a\, l^b\, l^c\, k^d \, ,
\hspace{11mm} \Psi_{2^{ijkl}}= C_{abcd}\; m_i^a\, m_j^b\, m_k^c\, m_l^d \, ,\nonumber \\
\Psi_{2T^{ij}}\rovno C_{abcd}\; k^a\, m_i^b\, l^c\, m_j^d \, ,
\hspace{10.2mm} \Psi_{2^{ij}}  = C_{abcd}\; k^a\, l^b\, m_i^c\, m_j^d \, , \label{defPsiCoef} \\
\Psi_{3T^{i}} \rovno C_{abcd}\; l^a\, k^b\, l^c\, m_i^d \, ,
\hspace{10.9mm} \Psi_{3^{ijk}} = C_{abcd}\; l^a\, m_i^b\, m_j^c\, m_k^d \, ,\nonumber\\
\Psi_{4^{ij}} \rovno C_{abcd}\; l^a\, m_i^b\, l^c\, m_j^d \, ,\nonumber
\end{eqnarray}
with indices ${\,i,j,k,l=2,\ldots,D-1\,}$, are components of the Weyl tensor with respect to the null frame ${\{\boldk, \boldl, \boldm_{i} \}}$ associated with ${\{\bolde_a \}}$ via relations ${\boldk=\ssqrt(\boldu+\bolde_{(1)})}$, ${\boldl=\ssqrt(\boldu-\bolde_{(1)})}$, ${\boldm_{i}=\bolde_{(i)}}$, see figure~\ref{figure1}.

The Weyl tensor components (\ref{defPsiCoef}) are listed by their boost weight and directly generalize the standard Newman--Penrose complex scalars $\Psi_A$ known from the ${D=4}$ case~\cite{KrtousPodolsky:2006,PodolskySvarc:2012}. In equations (\ref{invariant form of eq of geodesic deviationL}), (\ref{invariant form of eq of geodesic deviationT}), only the ``electric part'' of the Weyl tensor represented by the scalars in the left column of (\ref{defPsiCoef}) occurs. All these scalars respect the standard symmetries of the Weyl tensor, for example
\begin{equation}
\Psi_{4^{ij}} = \Psi_{4^{(ij)}} \, ,\hspace{2.4mm}  \Psi_{4^{k}}{}^{_k} = 0 \, ,\hspace{9.85mm}
\Psi_{3^{ijk}} = \Psi_{3^{i[jk]}} \, .\label{symmetries}
\end{equation}
Moreover, there are relations between the left and right columns of (\ref{defPsiCoef}), namely
\begin{eqnarray}
\Psi_{1T^i} \rovno \Psi_{1^{k}}{}^{_k}{}_{^i} \,,\quad
\Psi_{3T^i} = \Psi_{3^{k}}{}^{_k}{}_{^i} \,,\nonumber\\
\Psi_{2S} \rovno\Psi_{2T^{k}}{}^{_k}\,,\quad
\Psi_{2T^{[ij]}}={\textstyle\frac{1}{2}}\Psi_{2^{ij}} \, ,\quad \Psi_{2T^{(ij)}}={\textstyle\frac{1}{2}}\Psi_{2^{ikj}}{}^{_k}\,.\label{constraints}
\end{eqnarray}
Finally, let us remark that our notation which uses the symbols $\Psi_{A^{...}}$ in any dimension is related to the notations employed elsewhere, namely in  \cite{ColeyMilsonPravdaPravdova:2004,Coley:2008}, \cite{PravdaPravdovaColeyMilson:2004,PraPraOrt07}, and \cite{DurPraPraReall10,OrtaggioPravdaPravdova:2013}. Identifications for the components present in the invariant form of the equation of geodesic deviation are summarized in table~\ref{notationcomp} (more details can be found in \cite{PodolskySvarc:2012}).

\vspace{0.05cm}
\begin{table}[h]
\begin{tabular}{ccccccl}
\hline\hline
   $\Psi_{0^{ij}}$  &
   $\Psi_{1T^{j}}$  &
   $\Psi_{2S}$      &
   $\Psi_{2T^{ij}}$ &
   $\Psi_{3T^{j}}$  &
   $\Psi_{4^{ij}}$  &\\
\hline
   $ C_{0i0j}$  &
   $-C_{010j}$  &
   $-C_{0101}$  &
   $-C_{0i1j}$  &
   $ C_{101j}$  &
   $ C_{1i1j}$  &
   Coley \emph{et al.}~\cite{ColeyMilsonPravdaPravdova:2004,Coley:2008}\\
           &
   &
   $-\Phi$ &
   $-\Phi_{ij}$  &
   $\Psi_j$&
   $2\,\Psi_{ij}$&
   Pravda \emph{et al.}~\cite{PravdaPravdovaColeyMilson:2004,PraPraOrt07}\\
   $\Omega_{ij}$ &
   $-\Psi_j$     &
   $-\Phi$       &
   $-\Phi_{ij}$  &
   $\Psi'_j$     &
   $\Omega'_{ij}$&
   Durkee \emph{et al.}~\cite{DurPraPraReall10,OrtaggioPravdaPravdova:2013} \\
\hline\hline
\end{tabular}
\caption{\label{notationcomp}Different equivalent notations used in the literature for those Weyl scalars that occur in the equation of geodesic deviation (\ref{invariant form of eq of geodesic deviationL}), (\ref{invariant form of eq of geodesic deviationT}).}
\end{table}

\noindent
The remaining (independent) components of the Weyl tensor are listed in table~\ref{notationcomp2}.
\vspace{0.05cm}
\begin{table}[h]
\begin{tabular}{ccccccl}
\hline\hline
   $\Psi_{1^{ijk}}$  &
   $\Psi_{2^{ijkl}}$      &
   $\Psi_{2^{ij}}$ &
   $\Psi_{2T^{[ij]}}$  &
   $\Psi_{2T^{(ij)}}$  &
   $\Psi_{3^{ijk}}$  &\\
\hline
   $ C_{0ijk}$  &
   $ C_{ijkl}$  &
   $-C_{01ij}$  &
     &
     &
   $-C_{1ijk}$  &
   Coley \emph{et al.}~\cite{ColeyMilsonPravdaPravdova:2004,Coley:2008}\\
   $\Psi_{ijk}$ &
   $\Phi_{ijkl}$   &
   $-2\,\Phi^{\rm A}_{ij}$  &
   $-\Phi^{\rm A}_{ij}$     &
   $-\Phi^{\rm S}_{ij}$&
   $-\Psi'_{ijk}$  &
   Durkee \emph{et al.}~\cite{DurPraPraReall10,OrtaggioPravdaPravdova:2013} \\
\hline\hline
\end{tabular}
\caption{\label{notationcomp2} Other Weyl tensor components and their form in the GHP notation.}
\end{table}

For the most general Kundt spacetime (\ref{obecny Kundtuv prostorocas}), the \emph{null interpretation frame adapted to an arbitrary observer} moving along the timelike geodesic $\gamma(\tau)$, whose velocity vector  is ${\boldu=\dot{r}\,\mathbf{\partial}_{r}+\dot{u}\,\mathbf{\partial}_{u}+\dot{x}^p\mathbf{\partial}_{p}\,}$ (normalized as ${\boldu \cdot \boldu=-1}$, so that ${\dot{u}\not=0}$), takes the form\footnote{In this paper, ${i,j,k,l}$ are frame labels, whereas the indices ${p,q,m,n}$ denote the spatial coordinate components. For example, $m_{i}^{p}$ stands for the $p^{\hbox{\tiny{th}}}$  spatial coordinate component of the vector $\boldm_i$.}
\begin{eqnarray}
\boldk \rovno \frac{1}{\sqrt{2}\,\dot{u}}\,\mathbf{\partial}_r \, , \nonumber \\
\boldl\rovno \Big(\sqrt{2}\,\dot{r}-\frac{1}{\sqrt{2}\,\dot{u}}\Big)\mathbf{\partial}_r+\sqrt{2}\,\dot{u}\,\mathbf{\partial}_u+\sqrt{2}\,\dot{x}^p\mathbf{\partial}_p \, , \label{Kundt interpretation frame} \\
\boldm_i\rovno \frac{1}{\dot{u}}\,m_{i}^{p}(\,g_{pu}\,\dot{u}+g_{pq}\,\dot{x}^q)\,{\partial}_r+m_i^p\,\mathbf{\partial}_p \, , \nonumber
\end{eqnarray}
where ${\,m_i^p\,}$ satisfy ${{\,g_{pq}\,m_i^p \,m_j^q}=\delta_{ij}}$  to fulfil the normalization conditions ${\boldm_i \cdot \boldm_j=\delta_{ij}}$,  ${\boldk \cdot \boldl=-1}$. Notice that the vector $\boldk$ is oriented along the non-expanding, non-twisting and shear-free null congruence ${\mathbf{k}=\partial_r}$ defining the Kundt family. Moreover, ${\boldu=\ssqrt{(\boldk+\boldl)}}$ and ${\bolde_{(1)}=\ssqrt{(\boldk-\boldl)}=\sqrt{2}\,\boldk-\boldu}$. The spatial vector ${\bolde_{(1)}}$ is thus uniquely determined by the optically privileged null congruence of the Kundt family and the observer's velocity $\boldu$. For this reason we call such a special direction ${\bolde_{(1)}}$ \emph{longitudinal}, while the ${D-2}$ directions ${\bolde_{(i)}=\boldm_{i}}$ are \emph{transverse}.

To evaluate the scalars (\ref{defPsiCoef}) we have to project coordinate components of the Weyl tensor $C_{abcd}$ of the generic Kundt spacetime, which can be found in appendix~A of \cite{PodolskySvarc:2013}, onto the interpretation frame (\ref{Kundt interpretation frame}). Since ${C_{rprq}=0}$, we immediately obtain ${\Psi_{0^{ij}} \,= 0 }$, while the remaining Weyl scalars are in general non-zero. The relative motion of free test particles in any $D$-dimensional Kundt spacetime (\ref{obecny Kundtuv prostorocas}), determined by the equation of geodesic deviation (\ref{invariant form of eq of geodesic deviationL}), (\ref{invariant form of eq of geodesic deviationT}), can thus be naturally decomposed into the following components:

\begin{itemize}

\item
The presence of the \emph{cosmological constant} $\Lambda$ is encoded in the term
\begin{equation}\label{Lambda}
\left(
  \begin{array}{c}
    \ddot{Z}^{(1)} \\
    \ddot{Z}^{(i)} \\
  \end{array}
\right)
 = \frac{2\Lambda}{(D-1)(D-2)}
\left(
  \begin{array}{cc}
    1 & 0 \\
    0 & \delta_{ij} \\
  \end{array}
\right) \left(
  \begin{array}{c}
    Z^{(1)} \\
    Z^{(j)} \\
  \end{array}
\right) .
\end{equation}
It causes \emph{isotropic motion} of test particles, which is typical for (maximally symmetric) ``background'' spacetimes of constant curvature, namely Minkowski space, de~Sitter space and anti-de~Sitter space.

\item
The terms $\Psi_{2S}$ and $\Psi_{2T^{(ij)}}$ are responsible for \emph{Newtonian-like tidal deformations} since the motion of test particles is given by
\begin{equation}\label{Psi2e}
\left(
  \begin{array}{c}
    \ddot{Z}^{(1)} \\
    \ddot{Z}^{(i)} \\
  \end{array}
\right)
 =
\left(
  \begin{array}{cc}
    \Psi_{2S} & 0 \\
    0 & -\Psi_{2T^{(ij)}} \\
  \end{array}
\right) \left(
  \begin{array}{c}
    Z^{(1)} \\
    Z^{(j)} \\
  \end{array}
\right) ,
\end{equation}
where ${\Psi_{2S}=\Psi_{2T^{k}}{}^{_k} }$. These effects are typically present in type~D spacetimes.

\item
The scalars $\Psi_{3T^{j}}$ and $\Psi_{1T^{j}}$ represent \emph{longitudinal components} of the gravitational field resulting in specific motion associated with the spatial direction ${+\bolde_{(1)}}$ and ${-\bolde_{(1)}}$, respectively. Such terms cause accelerations
\begin{equation}\label{Psi3e}
\left(\!
  \begin{array}{c}
    \ddot{Z}^{(1)} \\
    \ddot{Z}^{(i)} \\
  \end{array}
\!\right)
 = -\frac{1}{\sqrt{2}}
\left(\!
  \begin{array}{cc}
    0 & \Psi_{{{\mathrm A}T}^{j}} \\
    \Psi_{{{\mathrm A}T}^{i}} & 0 \\
  \end{array}
\right) \left(
  \begin{array}{c}
    Z^{(1)} \\
    Z^{(j)} \\
  \end{array}
\right) , \nonumber
\end{equation}
where $\Psi_{\mathrm{A}T^{j}}$ stand for either $\Psi_{3T^{j}}$ or $-\Psi_{1T^{j}}$ which are mutually equivalent under ${\boldk \leftrightarrow \boldl}$, see (\ref{defPsiCoef}). These scalars combine motion in the privileged longitudinal  direction~$\pm\bolde_{(1)}$ with motion in the transverse spatial directions $\bolde_{(i)}$. Such effects typically occur in spacetimes of type~III, in particular III$_i$.

\item
The scalars $\Psi_{4^{ij}}$, characteristic for type~N spacetimes, can be interpreted as amplitudes of \emph{transverse gravitational waves} propagating along the spatial direction ${+\bolde_{(1)}}$. These components of the field influence test particles as
\begin{equation}\label{Psi4e}
\left(\!
  \begin{array}{c}
    \ddot{Z}^{(1)} \\
    \ddot{Z}^{(i)} \\
  \end{array}
\right)
 = -\frac{1}{2}
\left(
  \begin{array}{cc}
    0 & 0 \\
    0 & \Psi_{4^{ij}} \\
  \end{array}
\right) \left(
  \begin{array}{c}
    Z^{(1)} \\
    Z^{(j)} \\
  \end{array}
\right) .\nonumber
\end{equation}
They obviously cause purely transverse effects because there is no acceleration in the privileged longitudinal direction $\bolde_{(1)}$. The scalars $\Psi_{4^{ij}}$ form a symmetric traceless matrix of dimension ${(D-2)\times (D-2)}$, see (\ref{symmetries}). This matrix describing the amplitudes of gravitational waves has ${\frac12 D(D-3)}$ independent components corresponding to distinct polarization modes.

\end{itemize}

\noindent
More details about these general effects of any gravitational field on test particles can be found in \cite{PodolskySvarc:2012}. Their explicit discussion in the context of Kundt spacetimes will be presented in subsequent sections of this contribution.

There are also specific \emph{direct} effects of \emph{matter} in the equation of geodesic deviation (\ref{invariant form of eq of geodesic deviationL}), (\ref{invariant form of eq of geodesic deviationT}) which are determined by the frame components of the corresponding energy-momentum tensor $T_{ab}$. As an explicit illustration, we present here two physically interesting examples:

\begin{itemize}
\item
For \emph{pure radiation} (or ``null dust'') aligned along the null direction $\mathbf{k}$, the energy-momentum tensor  is
${T_{ab}=\rho\,{\mathrm k}_{a}{\mathrm k}_{b}}$ where $\rho$ represents radiation density. Its trace vanishes, ${T=0}$, and the only nonvanishing components of ${T_{ab}}$ in the equation of geodesic deviation (\ref{invariant form of eq of geodesic deviationL}), (\ref{invariant form of eq of geodesic deviationT}) reduce to
\begin{equation}\label{pure}
\left(
  \begin{array}{c}
    \ddot{Z}^{(1)} \\
    \ddot{Z}^{(i)} \\
  \end{array}
\right)
 = -\frac{4\pi\,\rho}{D-2}\,
\left(
  \begin{array}{cc}
    0 & 0 \\
    0 & \delta_{ij} \\
  \end{array}
\right) \left(
  \begin{array}{c}
    Z^{(1)} \\
    Z^{(j)} \\
  \end{array}
\right).
\end{equation}
There is no acceleration in the longitudinal spatial direction $\bolde_{(1)}$ and the effect in the transverse space is isotropic. Moreover, since ${\rho>0}$, it causes a radial contraction which may eventually lead to exact focusing.

\item
For an \emph{electromagnetic field aligned} with the Kundt geometry (i.e., ${{\cal F}_{ab}\,{\mathrm k}^b={\cal E}\, {\mathrm k}_a}$ where ${\mathbf{k}=\partial_r}$) the most general form of the Maxwell tensor is
\begin{equation}
{\cal F}={\cal E}\, \dd r\wedge \dd u+\sigma_p\, \dd u\wedge \dd x^p
+{\textstyle\frac{1}{2}}{\cal B}_{pq}\,\dd x^p\wedge \dd x^q\,.
\label{Maxwelltensor}
\end{equation}
Evaluating the energy-momentum tensor ${T_{ab}=\frac{1}{4\pi}\left({\cal F}_{ac}\,{{\cal F}_b}^{c}-\frac{1}{4}\,g_{ab}\,{\cal F}_{cd}\,{\cal F}^{cd}\right)}$ in the interpretation orthonormal frame, the corresponding effects on (uncharged) test particles, as given by the equation of geodesic deviation, take the form
\begin{equation}\label{elmag}
\left(
  \begin{array}{c}
    \ddot{Z}^{(1)} \\
    \ddot{Z}^{(i)} \\
  \end{array}
\right)
 =
\left(
  \begin{array}{cc}
    {\cal T}   & {\cal T}_j \\
    {\cal T}_i & {\cal T}_{ij} \\
  \end{array}
\right) \left(
  \begin{array}{c}
    Z^{(1)} \\
    Z^{(j)} \\
  \end{array}
\right),
\end{equation}
where
\begin{eqnarray}
&&\hspace{-20mm}
{\cal T}      =  -\frac{2}{D-2}\,\Big[({\cal E}^2+{\cal B}^2)+\frac{D-4}{D-1}({\cal E}^2-{\cal B}^2)\Big]\,, \nonumber \\
&&\hspace{-20mm}
{\cal T}_i    =  -\frac{2}{D-2}\,m_{i}^{p}\,\Big[\,\dot{x}^q ({\cal E}^2g_{pq}+{{\cal B}^2}_{\!\!pq})+\dot{u}\,({\cal E} {\cal E}_p-{\cal B}_{pq} {\cal E}_m\, g^{qm})\Big]\,, \\
&&\hspace{-20mm}
{\cal T}_{ij} =  \frac{2}{D-2}\,\Big[\,m_{i}^{p}m_{j}^{q}\,{{\cal B}^2}_{\!\!pq}-\delta_{ij}\Big({\cal B}^2+\frac{D-4}{D-1}({\cal E}^2-{\cal B}^2)\nonumber\\
&&\hspace{-10mm}+\dot{x}^p\dot{x}^q ({\cal E}^2g_{pq}+{\cal B}^2_{\,pq})+2\,\dot{u}\,\dot{x}^p ({\cal E} {\cal E}_p-{\cal B}_{pq} {\cal E}_m\, g^{qm})+\dot{u}^2\, {\cal E}_p {\cal E}_q\, g^{pq}\Big)\Big]\,, \nonumber
\end{eqnarray}
with convenient auxiliary variables defined as
\begin{equation}
\hspace{-20mm}
{\cal E}_p \equiv {\cal E} \,g_{up}-\sigma_p\,,\quad
{\cal B}^2_{\,pq} \equiv {\cal B}_{pm}{\cal B}_{qn}\,g^{mn}\,, \quad
{\cal B}^2 \equiv {\textstyle\frac{1}{2}}{\cal B}_{pm}{\cal B}_{qn}\,g^{pq}g^{mn}\,. \label{defB2}
\end{equation}
The motion simplifies considerably if the magnetic field is absent (${{\cal B}_{pq}=0}$):
\begin{eqnarray}
&&\hspace{-20mm}
{\cal T}      =  -\frac{2}{D-2}\,\frac{2D-5}{D-1}\,{\cal E}^2\,, \nonumber \\
&&\hspace{-20mm}
{\cal T}_i    =  -\frac{2}{D-2}\,m_{i}^{p}\Big(g_{pq}\,\dot{x}^q {\cal E}^2+\dot{u}\,{\cal E} {\cal E}_p\Big)\,, \\
&&\hspace{-20mm}
{\cal T}_{ij} =  -\frac{2}{D-2}\,\delta_{ij}\Big(\,\frac{D-4}{D-1}\,{\cal E}^2+g_{pq}\,\dot{x}^p\dot{x}^q {\cal E}^2+2\,\dot{u}\,\dot{x}^p {\cal E}{\cal E}_p+\dot{u}^2{\cal E}_p {\cal E}_q g^{pq} \, \Big)\,, \nonumber
\end{eqnarray}
in particular when ${D=4}$ and ${\sigma_p=0}$ (in which case ${{\cal E}_p = {\cal E} g_{up}}$).
\end{itemize}

\section{Explicit evaluation of the Weyl scalars $\Psi_{A^{...}}$}
\label{specfrmes}

The invariant amplitudes $\Psi_{A^{...}}$ of various gravitational field components (\ref{Psi2e})--(\ref{Psi4e}) combine the local \emph{curvature} of the Kundt spacetime with the \emph{kinematics} of  specific  motion along an arbitrary timelike geodesic~$\gamma(\tau)$. These should be evaluated at any given event corresponding to the actual position of the observer along $\gamma(\tau)$, with its actual velocity ${\boldu=\dot{r}\,\mathbf{\partial}_{r}+\dot{u}\,\mathbf{\partial}_{u}+\dot{x}^p\mathbf{\partial}_{p}\,}$.

The scalars $\Psi_{2S}$, $\Psi_{2T^{ij}}$, $\Psi_{1T^j}$, $\Psi_{3T^j}$, $\Psi_{4^{ij}}$ (and ${\Psi_{0^{ij}}=0}$) which enter the geodesic deviation equation  (\ref{invariant form of eq of geodesic deviationL}), (\ref{invariant form of eq of geodesic deviationT}) can most conveniently be expressed explicitly if we employ the relation between the \emph{interpretation null frame} (\ref{Kundt interpretation frame}) (adapted to the chosen geodesic observer) and the \emph{natural null frame} for the Kundt geometry (\ref{obecny Kundtuv prostorocas}) which is
\begin{eqnarray}
\boldk^\nat \rovno \mathbf{\partial}_r \, , \nonumber \\
\boldl^\nat \rovno {\textstyle\frac{1}{2}}g_{uu}\,\mathbf{\partial}_r+\mathbf{\partial}_u\, , \label{Kundt natural frame}\\
\boldm_i^\nat \rovno m_{i}^{p}\,(\,g_{up}\,{\partial}_r+\mathbf{\partial}_p) \, . \nonumber
\end{eqnarray}
The transition between the null frames (\ref{Kundt natural frame}) and (\ref{Kundt interpretation frame}) is a Lorentz transformation associated with the choice of different (timelike) observers, as explained in more detail in section~V and appendix~C of our work~\cite{PodolskySvarc:2012}. Specifically, the general interpretation frame is obtained from the natural one by combining a boost followed by a null rotation with fixed $\boldk^\nat$ (see equations (C3) and (C1) of~\cite{PodolskySvarc:2012}),
\begin{eqnarray}
\boldk   \rovno B\boldk^\nat\, , \nonumber \\
\boldl   \rovno B^{-1}\boldl^\nat + \sqrt2\,L^i \boldm_i^\nat + |L|^2 B\boldk^\nat\, , \label{Lorentztransf}\\
\boldm_i \rovno \boldm_i^\nat + \sqrt2\,L_i \,B\boldk^\nat\, , \nonumber
\end{eqnarray}
where ${|L|^2\equiv \delta^{ij}L_iL_j}$ and
\begin{equation}
B=\frac{1}{\sqrt2\,\dot{u}}\,, \qquad L_i=g_{pq}\,m_{i}^{p}\,\dot{x}^q\,.
\label{specialchoiceofframe}
\end{equation}
Conversely, the natural frame (\ref{Kundt natural frame}) is obtained from the interpretation frame (\ref{Kundt interpretation frame}) as a particular case when ${\,\sqrt{2}\,\dot{u}=1,\, \dot{x}^p=0\,}$ (and thus ${\,\sqrt{2}\,\dot{r}-1=\frac{1}{2}g_{uu}}$ due to the assumed normalization ${\boldu \cdot \boldu=-1}$), i.e., ${B=1,\, L_i=0}$. This corresponds to special observers with \emph{no motion in the transverse spatial directions} (${\,\dot{x}^p= 0\,}$ for all ${\,p=2,\ldots,D-1\,}$).

Under the Lorentz transformation (\ref{Lorentztransf}), the Weyl scalars change as
\begin{eqnarray}\label{Weylfixedk}
&&\hspace{-25mm}
{\Psi}_{0^{ij}}= 0 \,, \qquad
{\Psi}_{1T^{i}} = B\,\Psi_{1T^{i}}^\nat \,, \qquad
{\Psi}_{1^{ijk}}= B\,\Psi_{1^{ijk}}^\nat \,, \nonumber \\
&&\hspace{-25mm}
{\Psi}_{2S} = \Psi_{2S}^\nat-2\sqrt{2}\,\Psi_{1T^{i}}^\nat BL^i \,, \nonumber \\
&&\hspace{-25mm}
{\Psi}_{2^{ijkl}} = \Psi_{2^{ijkl}}^\nat-2\sqrt{2}\,B\!\left(L_{^{[}l}\Psi_{1^{k]ij}}^\nat-L_{^{[}i}\Psi_{1^{j]kl}}^\nat\right) \,, \nonumber \\
&&\hspace{-25mm}
{\Psi}_{2^{ij}} = \Psi_{2^{ij}}^\nat +\sqrt{2}\,\Psi_{1^{kij}}^\nat BL^k -2\sqrt{2}\,\Psi_{1T^{[i}}^\nat BL_{j^{]}} \,, \nonumber \\
&&\hspace{-25mm}
{\Psi}_{2T^{ij}} = \Psi_{2T^{ij}}^\nat+\sqrt{2}\,\Psi_{1^{ikj}}^\nat BL^k-\sqrt{2}\,\Psi_{1T^{i}}^\nat BL_j \,, \label{Psi_inter_nat}\\
&&\hspace{-25mm}
{\Psi}_{3^{ijk}}= B^{-1}\Psi_{3^{ijk}}^\nat +\sqrt{2}\left(\Psi_{2^{lijk}}^\nat L^l -\Psi_{2^{jk}}^\nat L_i +2L_{^{[}j}\Psi_{2T^{k]i}}^\nat\right) \nonumber \\
&&\hspace{-15mm}
  + 4\Psi_{1T^{[j}}^\nat L_{k^{]}}BL_i -2\left(\Psi_{1^{jli}}^\nat L_k +\Psi_{1^{ljk}}^\nat L_i -\Psi_{1^{kli}}^\nat L_j\right)BL^l + \Psi_{1^{ijk}}^\nat B|L|^2   \,, \nonumber \\
&&\hspace{-25mm}
{\Psi}_{3T^{i}} = B^{-1}\Psi_{3T^{i}}^\nat +\sqrt{2}\,\Psi_{2^{ij}}^\nat L^j -\sqrt{2}\left(\Psi_{2T^{ki}}^\nat L^k+\Psi_{2S}^\nat L_i\right) \nonumber \\
&&\hspace{-15mm}
  + 2\left(2\Psi_{1T^{j}}^\nat L_i -\Psi_{1^{kji}}^\nat L^k\right)BL^j -\Psi_{1T^{i}}^\nat B|L|^2 \,, \nonumber\\
&&\hspace{-25mm}
{\Psi}_{4^{ij}} = B^{-2}\Psi_{4^{ij}}^\nat+2\sqrt{2}\,B^{-1}\!\left(\Psi_{3T^{(i}}^\nat L_{j^{)}}-\Psi_{3^{(ij)k}}^\nat L^k\right) \nonumber \\
&&\hspace{-15mm}
  + 2\Psi_{2^{ikjl}}^\nat L^kL^l -4\Psi_{2T^{k(i}}^\nat L_{j^{)}}L^k +2\Psi_{2T^{(ij)}}^\nat |L|^2  - 2\Psi_{2S}^\nat L_iL_j -4\Psi_{2^{k(i}}^\nat L_{j^{)}}L^k \nonumber \\
&&\hspace{-15mm}
  - 2\sqrt{2}\,B\!\left(2\Psi_{1^{kl(i}}^\nat L_{j^{)}}L^kL^l +\Psi_{1^{(ij)k}}^\nat L^k|L|^2  + \Psi_{1T^{(i}}^\nat L_{j^{)}}|L|^2 -2\Psi_{1T^{k}}^\nat L^kL_iL_j\right)\!, \nonumber
\end{eqnarray}
see expressions (C5) and (C7) of~\cite{PodolskySvarc:2012} in the particular case when ${\Psi_{0^{ij}}=0}$. The scalars $\Psi_{A^{...}}^\nat$ represent the components (\ref{defPsiCoef}) of the Weyl tensor in the natural null frame (\ref{Kundt natural frame}). Recall that the coordinate components of $C_{abcd}$ were presented in appendix~A of \cite{PodolskySvarc:2013}. These scalars can also be used \emph{purely locally}. For some purposes, it is not necessary to evaluate all the functions along $\gamma(\tau)$ and express them in terms of the proper time $\tau$ of the geodesic observer. For example, to determine the \emph{algebraic type of the spacetime at any given event}, we only need to consider the values of the \emph{constants} $\Psi_{A^{...}}^\nat$ and their mutual relations. Moreover, they directly determine the \emph{actual acceleration} of test particles in various spatial directions.

\section{Algebraically special Kundt spacetimes}
\label{claasifKundt}

In our recent work~\cite{PodolskySvarc:2013}, we analyzed the geometric and algebraic properties of all Kundt spacetimes for which the optically privileged (non-expanding, non-twisting, shear-free)  congruence is generated by the null vector field ${\mathbf{k}=\partial_r}$ that is a \emph{multiple WAND} (Weyl aligned null direction).

Specifically, ${\Psi_{0^{ij}}^\nat \,= 0 }$ immediately confirms the results of \cite{OrtaggioPravdaPravdova:2007,PodolskyZofka:2009,OrtaggioPravdaPravdova:2013} that a generic  Kundt geometry represented by the metric (\ref{obecny Kundtuv prostorocas}) is (at least) of algebraic type~I (subtype~I(b), in fact) and ${\mathbf{k}=\boldk^\nat}$ is~WAND. In~\cite{PodolskySvarc:2013} we also demonstrated that
\emph{the general Kundt spacetime of algebraic type}~II \emph{with respect to the double WAND}~$\mathbf{k}$ in any dimension $D$ can be written in the form (\ref{obecny Kundtuv prostorocas}) with ${g_{up}=e_p+ f_p \,r}$ \emph{at most linear} in~$r$,
\begin{equation}
\hspace{-15mm}
\dd s^2 = g_{pq} \,\dd x^p\dd x^q+2\,(e_p+ f_p \,r)\,\dd u\,\dd x^p -2\,\dd u\,\dd r+g_{uu}(r,u,x)\,\dd u^2 \, , \label{obecny Kundt II}
\end{equation}
where $g_{pq}(u,x)$, $e_p(u,x)$, $f_p(u,x)$,  ${p=2,\ldots,D-1}$, are functions independent of~$r$.

For such algebraically special Kundt geometries (\ref{obecny Kundt II}) with the multiple WAND ${\mathbf{k}=\boldk^\nat}$ there is ${\Psi_{1T^j}^\nat=0={\Psi}_{1^{ijk}}^\nat}$. Moreover, in~\cite{PodolskySvarc:2013} we \emph{explicitly} evaluated all the remaining Weyl scalars $\Psi_{A^{...}}^\nat$ of the boost weights ${0,-1,-2}$. After
lengthy calculations, we obtained the following surprisingly simple expressions, namely
\begin{eqnarray}
&&\hspace{-20mm}
\Psi_{2S}^\nat = \frac{D-3}{D-1}\,\Big[\,\frac{1}{2}\,g_{uu,rr}-\frac{1}{4}f^p f_p+\frac{1}{D-2}\Big(\frac{\,^{S}\!R}{D-3}+f\Big)\Big], \label{Psi2s}\\
&&\hspace{-20mm}
\Psi_{2T^{(ij)}}^\nat = \tilde\Psi_{2T^{(ij)}}^\nat+\frac{1}{D-2}\,\delta_{ij}\,\Psi_{2S}^\nat, \label{Psi2T(ij)}\\
&&\hspace{-20mm}
\tilde\Psi_{2T^{(ij)}}^\nat = \frac{m_{i}^{p}m_{j}^{q} }{D-2}\,\Big[\Big({\!\,^{S}\!R_{pq}-\frac{1}{D-2}\,g_{pq}\,^{S}\!R\Big)
+\frac{1}{2}(D-4)\Big(f_{pq}-\frac{1}{D-2}\,g_{pq}\,f\Big)}\Big]\,, \label{tildePsi2T(ij)}\\
&&\hspace{-20mm}
\Psi_{2^{ijkl}}^\nat= \tilde \Psi_{2^{ijkl}}^\nat
-\frac{2}{(D-3)(D-4)}\,(\delta_{ik}\,\delta_{jl}-\delta_{il}\,\delta_{jk})\,\Psi_{2S}^\nat \nonumber\\
&&\hspace{1.4mm} +\frac{2}{D-4}\,(\delta_{ik}\Psi_{2T^{(jl)}}^\nat-\delta_{il}\Psi_{2T^{(jk)}}^\nat
-\delta_{jk}\Psi_{2T^{(il)}}^\nat+\delta_{jl}\Psi_{2T^{(ik)}}^\nat)\,,\label{Psi2ijkl}\\
&&\hspace{-20mm}
\tilde \Psi_{2^{ijkl}}^\nat= m_{i}^{m}m_{j}^{p}m_{k}^{n}m_{l}^{q}\,^{S}\!C_{mpnq} \,,\label{tildePsi2ijkl}\\
&&\hspace{-20mm}
\Psi_{2^{ij}}^\nat = 2\,\Psi_{2T^{[ij]}}^\nat = m_{i}^{p}m_{j}^{q}\,F_{pq}\,,\label{Psi2ij}\\
&&\hspace{-20mm}
\Psi_{3T^j}^\nat = -m_{j}^{p}\,\frac{D-3}{D-2}\,\Big[\,\frac{1}{2} \Big(rf_p\,g_{uu,rr}
+g_{uu,rp}-f_{p,u}\Big)+e_p\Big(\frac{1}{2}\,g_{uu,rr}-\frac{1}{4}f^qf_q\Big) \nonumber \\
&&\hspace{1.4mm}+\frac{1}{4}f^qe_qf_p-\frac{1}{2}f^qE_{qp}-\frac{1}{D-3}\,X_p
-r\,\Big(\frac{1}{2}f^q F_{qp}+\frac{1}{D-3}Y_p\Big)\Big],\label{Psi 3TjD}\\
&&\hspace{-20mm}
\Psi_{3^{ijk}}^\nat = \tilde\Psi_{3^{ijk}}^\nat +\frac{1}{D-3}\,(\delta_{ij}\,\Psi_{3T^k}^\nat-\delta_{ik}\,\Psi_{3T^j}^\nat)\, , \label{Psi3D}\\
&&\hspace{-20mm}
\tilde\Psi_{3^{ijk}}^\nat = m_{i}^{p}m_{j}^{m}m_{k}^{q}\,
\Big[\Big(X_{pmq}-\frac{2}{D-3}\, g_{p[m}\,X_{q]}\Big)+r\,\Big(Y_{pmq}-\frac{2}{D-3}\, g_{p[m}\,Y_{q]}\Big)\Big] \, , \label{Psi3ijkD}\\
&&\hspace{-20mm}
\Psi_{4^{ij}}^\nat =  m_{i}^{p}m_{j}^{q}\, \Big(W_{pq}-\frac{1}{D-2}\, g_{pq}\,W\Big)  \, , \label{Psi4D}
\end{eqnarray}
\vspace{1.3mm}
in which ${{\,g_{pq}\,m_i^p \,m_j^q}=\delta_{ij}\,}$,
\begin{eqnarray}
&&\hspace{-20.9mm}
X_{pmq} \equiv
e_{[q||m]||p}+F_{qm}\,e_p + F_{p[m}   \,e_{q]} + e_{p[m}f_{q]}-f_{p[m}\,e_{q]}+g_{p[m,u||q]}\, , \label{Xipj*} \\
&&\hspace{-20mm}
Y_{pmq} \equiv
f_{[q||m]||p}+ F_{qm}\,f_p + F_{p[m}\,f_{q]} , \label{YpmgD}\\
&&\hspace{-18.7mm}
W_{pq} \equiv {\textstyle
-\frac{1}{2}(g_{uu})_{||p||q}+\frac{1}{2}g_{uu}f_{(p||q)}+\frac{1}{2}g_{uu,(p}f_{q)}} \nonumber\\
&&\hspace{-8mm} {\textstyle
-\frac{1}{2}g_{uu,r}(r f_{(p||q)}+e_{pq})-\frac{1}{2}g_{uu,rr}(r^2 f_p f_q+2r f_{(p} e_{q)}+e_pe_q)}
\nonumber\\
&&\hspace{-8mm} {\textstyle
+\frac{1}{2}[(f_{p,u}-g_{uu,rp})(r f_q+e_q)+(f_{q,u}-g_{uu,rq})(r f_p+e_p)]}
\nonumber\\
&&\hspace{-8mm} {\textstyle
+ r^2 g^{mn}F_{mp}F_{nq} + r\big(f_{(p,u||q)}-2g^{mn}E_{m(p}F_{q)n} }
+ f^m F_{m(p}\,e_{q)}-e^m F_{m(p}\,f_{q)}\big)  \nonumber\\
&&\hspace{-8mm} {\textstyle
+ e_{(p,u||q)}-\frac{1}{2}g_{pq,uu}+g^{mn}E_{mp}E_{nq} }
+ f^mE_{m(p}\,e_{q)}-e^mE_{m(p}\,f_{q)} \nonumber\\
&&\hspace{-8mm} {\textstyle
+\frac{1}{4}(e^me_mf_pf_q+f^mf_me_pe_q)-\frac{1}{2}f^me_mf_{(p}e_{q)}  }
\, , \label{WpqD}
\end{eqnarray}
and their contractions are
\begin{equation}
X_q \equiv g^{pm}X_{pmq}\,,\quad Y_q \equiv g^{pm}\,Y_{pmq}\,, \quad W\equiv g^{pq}\,W_{pq}\,.\label{W}
\end{equation}
Note that ${\,W_{pq} = W_{qp}\,}$, while ${\,X_{pmq} = -X_{pqm}\,}$ and ${\,Y_{pmq} = -Y_{pqm}}$, so that ${X_q}$ and ${Y_q}$ are the only non-trivial contractions of $X_{pmq}$ and $Y_{pmq}$, respectively.

In these expressions we have introduced convenient geometric quantities
\begin{eqnarray}
f^p &\equiv& g^{pq}f_q\,,\label{defkontra}\\
f_{p||q} &\equiv& f_{p,q}-\Gamma^m_{pq} f_m \,, \label{defderiv}\\
{f^p}_{||p} &\equiv& g^{pq}f_{p||q}\,, \label{defdiv}\\
f_{pq} &\equiv& f_{(p||q)} + {\textstyle\frac{1}{2}} f_p f_q\,, \label{deffij}\\
f &\equiv& g^{pq}f_{pq}={f^p}_{||p}+ {\textstyle\frac{1}{2}} f^p f_p\,, \label{deff}\\
F_{pq} &\equiv&  f_{[p||q]} = f_{[p,q]} \,, \label{deffija}\\
f_{[m||q]||p} &\equiv& f_{[m,q],p}-\Gamma^{n}_{pm}f_{[n,q]}-\Gamma^{n}_{pq}f_{[m,n]}\,, \label{defidef1t}\\
f_{p,u||q} &\equiv& (f_{p,u})_{||q}= f_{p,uq}-f_{n,u}\,\Gamma^{n}_{pq}\,, \label{defideeu11t}\\
f_{(p,u||q)} &\equiv& f_{(p,q),u}-f_{n,u}\,\Gamma^{n}_{pq}\,, \label{defideeu1t}\\
e^p &\equiv& g^{pq}e_q\,,\label{defkontrae}\\
e_{p||q} &\equiv& e_{p,q}-\Gamma^m_{pq} e_m \,, \label{defderivapet}\\
e_{pq} &\equiv& e_{(p||q)} - {\textstyle\frac{1}{2}} g_{pq,u}\,, \label{defeijapt}\\
E_{pq} &\equiv& e_{[p||q]}\,+{\textstyle \frac{1}{2}}g_{pq,u}\,, \label{defEijapt}\\
e_{[m||q]||p} &\equiv& e_{[m,q],p}-\Gamma^{n}_{pm}e_{[n,q]}-\Gamma^{n}_{pq}e_{[m,n]}\,, \label{defidee1t}\\
e_{p,u||q} &\equiv& (e_{p,u})_{||q}= e_{p,uq}-e_{n,u}\,\Gamma^{n}_{pq}\,, \label{defideeu22t}\\
e_{(p,u||q)} &\equiv& e_{(p,q),u}-e_{n,u}\,\Gamma^{n}_{pq}\,, \label{defideeu2t}\\
g_{p[m,u||q]} &\equiv&  g_{p[m,q],u}+{\textstyle \frac{1}{2}}(\,\Gamma^{n}_{pm}\,g_{nq,u}-\Gamma^{n}_{pq}\,g_{nm,u}) \,,\label{defideeu28t}\\
(g_{uu})_{||p||q} &\equiv& g_{uu,pq}-g_{uu,n}\,\Gamma^{n}_{pq}\,, \label{defideeu3t}\\
\triangle g_{uu} &\equiv& g^{pq}(g_{uu})_{||p||q}\,. \label{Laplaceguut}
\end{eqnarray}
The symbol $||$ indicates covariant derivative with respect to the spatial metric~$g_{pq}$ in the transverse ${(D-2)}$-dimensional Riemannian space. The corresponding Riemann and Ricci tensors are ${\,^{S}\!R_{mpnq}}$ and ${\,^{S}\!R_{pq}}$, the Ricci scalar is ${\,^{S}\!R}$ and the Weyl tensor reads ${\,^{S}\!C_{mpnq}}$.
All the quantities (\ref{defkontra})--(\ref{defideeu28t}) are \emph{independent} of the coordinate~$r$.

\renewcommand{\arraystretch}{1.4}
\begin{table}[b]
\begin{tabular}{cc}
\hline\hline
  Type & Necessary and sufficient conditions \\
\hline\hline
II(a)  & ${g_{uu}=a(u,x)\,r^2+ b(u,x)\,r+c(u,x)}$ \ where\
${\,a=\frac{1}{4}f^p f_p-\frac{1}{D-2}\big(\frac{\,^{S}\!R}{D-3}+f\big)}$ \\
\hline
II(b)  & ${\,^{S}\!R_{pq}-\frac{1}{D-2}\,g_{pq}\,^{S}\!R
=-\frac{1}{2}(D-4)\big(f_{pq}-\frac{1}{D-2}\,g_{pq}\,f\big)}$ \\
\hline
II(c)  & ${\,^{S}\!C_{mpnq}=0}$  \\
\hline
II(d)  & ${F_{pq}=0}$  \\
\hline\hline
III & II(abcd)   \\
\hline
III(a) & ${ a_{,p}+f_{p}\,a = 0\,}$ \ where\  ${\,a=\frac{1}{4}f^p f_p-\frac{1}{D-2}\big(\frac{\,^{S}\!R}{D-3}+f\big)}$ \\
       & ${ \ \  b_{,p}-f_{p,u}=
       \frac{2}{D-2}\,e_{p}\big(\frac{\,^{S}\!R}{D-3}+f\big)-\frac{1}{2}f^qe_qf_p+f^q E_{qp}+\frac{2}{D-3}\,X_p}$  \\
\hline
III(b) & ${X_{pmq}=\frac{1}{D-3}\big( g_{pm}\,X_q-g_{pq}\,X_m\big)}$ \\
\hline\hline
N & III(ab)  \\
\hline\hline\hline
O & N with ${\,W_{pq}=\frac{1}{D-2}\, g_{pq}\,W}$  \ (special case O' is ${\,W_{pq}=0}$) \\
\hline\hline\hline
D & ${\frac{1}{2} \big(rf_p\,g_{uu,rr}+g_{uu,rp}-f_{p,u}\big)
+e_p\big(\frac{1}{2}\,g_{uu,rr}-\frac{1}{4}f^qf_q\big)\quad}$   \\[-1mm]
 & ${\qquad=r\,\big(\frac{1}{2}f^q F_{qp}+\frac{1}{D-3}Y_p\big)
-\frac{1}{4}f^qe_qf_p+\frac{1}{2}f^qE_{qp}+\frac{1}{D-3}\,X_p}$   \\
 & ${ X_{pmq}=\frac{1}{D-3}\big(\, g_{pm}\,X_q-g_{pq}\,X_m\big)}$ \  and \  ${ Y_{pmq} =\frac{1}{D-3}\big(\, g_{pm}\,Y_q-g_{pq}\,Y_m\big)}$ \\
  & ${ W_{pq}=\frac{1}{D-2}\, g_{pq}\,W  }$ \\
\hline\hline
\end{tabular}
\caption{\label{classifKundt} The classification scheme of algebraically special Kundt geometries (\ref{obecny Kundt II}) in any dimension~$D$ with ${\mathbf{k}=\boldk^\nat=\partial_r}$ being a multiple WAND. For type~D subclass, the vector ${\boldl^\nat=\frac{1}{2}g_{uu}\,\mathbf{\partial}_r+\mathbf{\partial}_u}$ is a double WAND.  If all conditions for type~D are satisfied \emph{and} conditions for the subtypes~II(a), II(d), II(c), II(d) are also valid, we obtain the subtypes D(a), D(b), D(c), D(d), respectively. The subtype D(abcd) is equivalent to type~O. In the classic ${D=4}$ case, conditions for II(b), II(c) and III(b) are always satisfied.}
\end{table}

The explicit Weyl scalars (\ref{Psi2s})--(\ref{Psi4D}) in the natural frame (\ref{Kundt natural frame}) enabled us in~\cite{PodolskySvarc:2013} to determine, without assuming any field equations, the \emph{classification scheme of all algebraic types and subtypes} with respect to the multiple WAND ${\mathbf{k}=\partial_r}$. Summary of such Kundt geometries (\ref{obecny Kundt II}) in any dimension~$D$ is presented in table~\ref{classifKundt}.

\section{Geodesic deviation in Kundt spacetimes with a multiple WAND ${\mathbf{k}}$}
\label{subclasses}

In the remaining parts of this paper we will discus an important family of algebraically special Kundt spacetimes (\ref{obecny Kundt II}). As described in the previous section, Weyl scalars of the two highest boost weights vanish identically,  ${\Psi_{0^{ij}}^\nat=0}$, ${\,\Psi_{1T^i}^\nat=0={\Psi}_{1^{ijk}}^\nat}$. From (\ref{Psi_inter_nat}) we then immediately obtain
\begin{equation}\label{Psi0a1}
\Psi_{0^{ij}} = 0 \, , \qquad \Psi_{1T^i} = 0\, , \qquad \Psi_{1^{ijk}} = 0 \, .
 \end{equation}
The geodesic deviation equations (\ref{invariant form of eq of geodesic deviationL}), (\ref{invariant form of eq of geodesic deviationT}) (omitting the frame components of $T_{ab}$ encoding the direct influence of matter, for example (\ref{pure}) or (\ref{elmag})) for the case of Kundt class of (vacuum) spacetimes (\ref{obecny Kundt II}) thus reduce to
\begin{eqnarray}
&&\hspace{-15mm}
\ddot{Z}^{(1)} = \frac{2\Lambda}{(D-2)(D-1)}\,Z^{(1)} +\Psi_{2S}\,Z^{(1)}
- \frac{1}{\sqrt{2}}\,\Psi_{3T^j}\,Z^{(j)} \, , \label{Kundt geodesic deviationA}\\
&&\hspace{-14.6mm}
\ddot{Z}^{(i)} = \frac{2\Lambda}{(D-2)(D-1)}\,Z^{(i)}-\Psi_{2T^{(ij)}}\,Z^{(j)}
- \frac{1}{\sqrt{2}}\,\Psi_{3T^i}\,Z^{(1)} -\frac{1}{2}\,\Psi_{4^{ij}}\,Z^{(j)}\, .\label{Kundt geodesic deviationB}
\end{eqnarray}
The corresponding Weyl scalars in the interpretation null frame are given by expressions (\ref{Psi_inter_nat}) with (\ref{specialchoiceofframe}), which now simplify considerably due to (\ref{Psi0a1}):
\begin{eqnarray}
&&\hspace{-20mm}
\Psi_{2S} = \Psi_{2S}^\nat \, , \qquad \Psi_{2T^{ij}} = \Psi_{2T^{ij}}^\nat\,, \nonumber\\
&&\hspace{-21.2mm}
\Psi_{3T^i} = \sqrt{2}\,\dot{u}\,\Psi_{3T^{i}}^\nat
+\sqrt{2}\,\dot{x}^p\,g_{pq}\,\Big(\big(\Psi_{2^{ij}}^\nat -\Psi_{2T^{ji}}^\nat\big) \,m^{jq}-\Psi_{2S}^\nat \,m_i^q\Big)\,, \label{Psinatn}\\
&&\hspace{-20.2mm}
\Psi_{4^{ij}} = 2\,\dot{u}^2\Psi_{4^{ij}}^\nat
+4\,\dot{u}\,\dot{x}^p\,g_{pq}\,\Big(\Psi_{3T^{(i}}^\nat m^q_{j^{)}}-\Psi_{3^{(ij)k}}^\nat m^{kq}\Big)
 \nonumber \\
&&\hspace{-10mm}
+2\,\dot{x}^p\dot{x}^q\Big(
\,g_{pq}\,\Psi_{2T^{(ij)}}^\nat -g_{pm}\,g_{qn}\,\Psi_{2S}^\nat\, m_i^m m_j^n +g_{pm}\,g_{qn}\,\Psi_{2^{ikjl}}^\nat m^{km} m^{ln}
 \nonumber \\
&& \hspace{4mm}
-2\,g_{pm}\,g_{qn}\,\big(\Psi_{2^{k(i}}^\nat +\Psi_{2T^{k(i}}^\nat \big)\,m^n_{j^{)}} m^{k m}\Big)
\, ,\nonumber
 \end{eqnarray}
where ${m^{jq}\equiv \delta^{ij}m_i^{\,q}}$ and both the frame indices ${\,i,j,k,l}$ and the coordinate indices ${p,q,m,n}$ take the ranges ${2,3,\ldots,D-1}$. The coefficients $\Psi_{A^{...}}^\nat$ are explicitly given by expressions (\ref{Psi2s})--(\ref{Psi4D}).

For completeness, the remaining Weyl tensor components in the interpretation frame that do not enter \emph{directly}  the equations of geodesic deviation  (\ref{Kundt geodesic deviationA}), (\ref{Kundt geodesic deviationB}) are
\begin{eqnarray}
&&\hspace{-16.8mm}
\Psi_{2^{ij}}= \Psi_{2^{ij}}^\nat\,, \qquad  \Psi_{2^{ijkl}}= \Psi_{2^{ijkl}}^\nat\,, \nonumber\\
&&\hspace{-18mm}
\Psi_{3^{ijk}}= \sqrt{2}\,\dot{u}\,\Psi_{3^{ijk}}^\nat
+\sqrt{2}\,\dot{x}^p\,g_{pq}\,\Big(\Psi_{2^{lijk}}^\nat \,m^{lq}-\Psi_{2^{jk}}^\nat \,m_i^q+2\,m_{^{[}j}^{\,q}\Psi_{2T^{k]i}}^\nat\Big)\,.\label{Psinatnext}
 \end{eqnarray}

The specific relative motion of free test particles in any algebraically special Kundt spacetime (\ref{obecny Kundt II}) with a multiple WAND ${\mathbf{k}}$ thus consists of \emph{isotropic influence} of the cosmological constant~$\Lambda$, \emph{Newtonian-like tidal deformations} represented by $\Psi_{2S}$, $\Psi_{2T^{(ij)}}$, \emph{longitudinal accelerations} associated with the direction $+\mathbf{e}_{(1)}$ given by $\Psi_{3T^j}$, and by \emph{transverse gravitational waves} propagating along $+\mathbf{e}_{(1)}$ encoded in the symmetric traceless matrix $\Psi_{4^{ij}}$. These components were described separately in (\ref{Lambda})--(\ref{Psi4e}). The invariant amplitudes (\ref{Psinatn}) combine the curvature of the Kundt spacetime with the kinematics of the specific geodesic motion. In contrast to longitudinal and transverse effects, the Newtonian-like deformations caused by $\Psi_{2S}$ and $\Psi_{2T^{(ij)}}$ are \emph{independent} of the observer's velocity components $\dot{x}^p$ and $\dot{u}$.

We will now describe systematically the canonical structure of relative motion of free test particles in all possible algebraic types and subtypes of the Kundt family summarized in table~\ref{classifKundt}.

\subsection{Type~O Kundt spacetimes}
\label{typO}

For type~O Kundt spacetimes, the Weyl tensor vanishes identically, so that \emph{all} the Weyl scalars $\Psi_{A^{...}}^\nat$ given by (\ref{Psi2s})--(\ref{Psi4D}) \emph{are zero}. In view of (\ref{Psinatn}), the geodesic deviation equations (\ref{Kundt geodesic deviationA}), (\ref{Kundt geodesic deviationB}) for the type~O \emph{vacuum} Kundt spacetimes reduce to
\begin{eqnarray}
&&\hspace{-20mm}   \ddot{Z}^{(1)} = \frac{2\Lambda}{(D-2)(D-1)}\,Z^{(1)} \,, \qquad
\ddot{Z}^{(i)} = \frac{2\Lambda}{(D-2)(D-1)}\,Z^{(i)} \,. \label{Kundt geodesic deviationB_typeO}
\end{eqnarray}
There is thus no distinction between the (generically privileged) longitudinal spatial direction~$\bolde_{(1)}$ and the transverse spatial directions $\bolde_{(i)}$, ${\,i=2,\ldots,D-1}$. The relative motion is \emph{isotropic} and fully determined by the cosmological constant $\Lambda$, see~(\ref{Lambda}). This is in full agreement with the well-known fact~\cite{Stephani:2003, GriffithsPodolsky:2009} that the only type~O vacuum spaces are just Minkowski space, de~Sitter space or anti-de~Sitter space.

For \emph{non-vacuum} type~O (conformally flat) Kundt spacetimes, it is necessary to add the terms representing direct influence of  matter. For example, in the case of \emph{pure radiation} (``null dust'') aligned along  $\mathbf{k}$, the components (\ref{pure}) have to be superposed, and the equations of geodesic deviation become
\begin{eqnarray}
&&\hspace{-20mm}
\ddot{Z}^{(1)} = \frac{2\Lambda}{(D-2)(D-1)}\,Z^{(1)}\,, \qquad
\ddot{Z}^{(i)} = \frac{2\Lambda}{(D-2)(D-1)}\,Z^{(i)} -\frac{4\pi\,\rho}{D-2}\,Z^{(i)}\,.
\label{Kundt geodesic deviationB_typeO_purad}
\end{eqnarray}
Since ${\rho>0}$, there is now an additional radial contraction in the transverse subspace.

For aligned electromagnetic field, the additional matter terms are given by (\ref{elmag}).

\subsection{Type~N Kundt spacetimes}
\label{typN}

As shown in~\cite{PodolskySvarc:2013}, for type~N Kundt spacetimes (\ref{obecny Kundt II}) (with quadruple WAND ${\mathbf{k}=\partial_r}$) the only non-trivial Weyl scalars are ${\Psi_{4^{ij}}^\nat}$. Considering (\ref{Psinatn}), the geodesic deviation equations (\ref{Kundt geodesic deviationA}), (\ref{Kundt geodesic deviationB}) for vacuum type~N spacetimes thus take the form
\begin{eqnarray}
&&\hspace{-1mm}
\ddot{Z}^{(1)} = \frac{2\Lambda}{(D-2)(D-1)}\,Z^{(1)} \, , \label{Kundt geodesic deviationA_typeN}\\
&&\hspace{-0.6mm}
\ddot{Z}^{(i)} = \frac{2\Lambda}{(D-2)(D-1)}\,Z^{(i)} -\dot{u}^2\,\Psi_{4^{ij}}^\nat\,Z^{(j)}\, ,\label{Kundt geodesic deviationB_typeN}
\end{eqnarray}
where, due to (\ref{Psi4D}),
\begin{equation}
\Psi_{4^{ij}}^\nat= m_{i}^{p}m_{j}^{q}\, \Big(W_{pq}-\frac{1}{D-2}\, g_{pq}\,W\Big)
\end{equation}
is a \emph{symmetric and traceless matrix} fully determined by $W_{pq}$. The symmetric matrix $W_{pq}$ introduced in (\ref{WpqD}) simplifies, using all relevant conditions in table~\ref{classifKundt} (cf.~\cite{PodolskySvarc:2013}), to
\begin{eqnarray}
&&\hspace{-25mm} W_{pq} \equiv {\textstyle r\,\Big[\,\frac{1}{2}\,a\, g_{pq,u}+U_{(p||q)}+U_{(p} f_{q)}\, \Big]} \nonumber\\
&&\hspace{-16mm}
{\textstyle -\frac{1}{4}\Big[\big(c_{,p}-c\,f_{p}\big)_{||q}+\big(c_{,q}-c\,f_{q}\big)_{||p}\Big]
-\frac{1}{2}b\,e_{pq} +\Big(a-\frac{1}{4}f^{m}f_{m}\Big)e_{p}e_{q}+Z_{(pq)}}\, , \label{WpqN}
\end{eqnarray}
in which
\begin{eqnarray}
&&\hspace{-23.8mm} U_p \equiv {\textstyle \frac{1}{2}f_{p,u}-\frac{1}{4}f^{q}f_{q}\,e_{p}+\frac{1}{4}f^{q}e_{q}f_{p}-\frac{1}{2}f^{q}E_{qp}-\frac{1}{D-3}\,X_{p}}\, , \label{defUp}\\
&&\hspace{-25mm} Z_{pq} \equiv
{\textstyle \frac{1}{4}e^{m}e_{m}\,f_{p}f_{q} +e_{p,u||q}-\frac{1}{2}\,g_{pq,uu}-e^{m}E_{mp}f_{q}
+g^{mn}E_{mp}E_{nq}-\frac{2}{D-3}\,X_{p}\,e_{q}} \, , \label{Zpq}
\end{eqnarray}
and its trace is ${W=g^{pq}W_{pq}\,}$. The matrix~${\Psi_{4^{ij}}^\nat}$ represents the \emph{amplitudes of Kundt gravitational waves} in any dimension $D$. In general, their effect is superposed on the isotropic influence of the cosmological constant $\Lambda$, as given by (\ref{Kundt geodesic deviationB_typeO}). In the case ${D=4}$ this was analyzed and described in our previous work~\cite{BicakPodolsky:1999b}.

\begin{figure}[t]
  \begin{center}
  \includegraphics[width=0.82\textwidth]{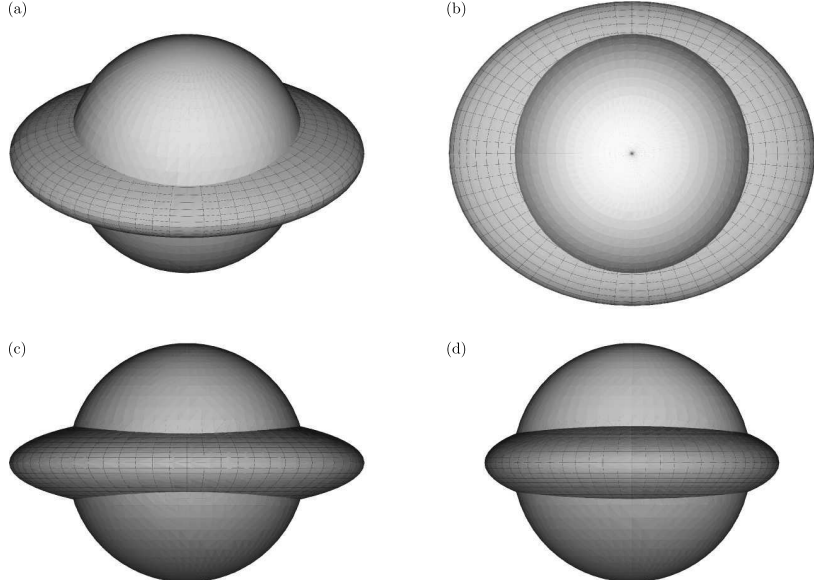}
  \end{center}
   \vspace{-3mm}
  \caption{\label{figure2} Deformation of a sphere of test particles in the case when two eigenvalues of ${-\Psi_{4^{ij}}^\nat}$ are positive and one is negative (${D=5}$, the wave propagates in the direction $\bolde_{(1)}$, the transverse 3-space shown is spanned by $\bolde_{(2)},\bolde_{(3)},\bolde_{(4)}$). Plot (a) is a global view, (b), (c), (d) are views from top, front, right, respectively.
}
\end{figure}

Since the set of ${(D-2)\times (D-2)}$  scalars ${\Psi_{4^{ij}}^\nat}$ forms a symmetric and traceless matrix, it has in general ${N\equiv{\frac{1}{2}D(D-3)}}$ independent components corresponding to the \emph{polarization modes} of the gravitational wave. The remaining freedom in the choice of the transverse vectors $\boldm_i$ of the interpretation frame (\ref{Kundt interpretation frame}) is given by spatial rotations ${{\boldm}'_{i}=\Phi_i{}^j\,\boldm_{j}}$, where ${\Phi_i{}^j\,\Phi_k{}^l\,\delta_{jl}=\delta_{ik}}$,
which leave the null frame vectors ${\boldk, \boldl}$ unchanged. These rotations belong to $\mathrm{SO}(D-2)$ group with ${N_{\mathrm{rot}}\equiv\frac{1}{2}(D-2)(D-3)}$ independent generators. Therefore, the number of physical degrees of freedom is
\begin{equation}
N-N_{\mathrm{rot}}=D-3 \ .
\end{equation}
This is exactly the \emph{number of independent eigenvalues} of the matrix $\Psi_{4^{ij}}^\nat$ which fully characterize the geodesic deviation deformation of a set of test particles. The sum of all these eigenvalues must vanish (the traceless property), so that there is at least one positive and one negative eigenvalue. The number of distinct options of dividing the remaining eigenvalues into three groups with positive, null and negative signs is ${D-2 \choose 2}$. Concerning the signs of the eigenvalues, we can thus distinguish ${\frac{1}{2}(D-2)(D-3)}$ \emph{geometrically and physically distinct cases}.

Diagonalizing ${-\Psi_{4^{ij}}^\nat}$ and denoting its eigenvalues as ${{\cal A}_2, {\cal A}_3, \ldots\,}$, we obtain
\begin{equation}\label{diagtrace}
-\Psi_{4^{ij}}^\nat=\hbox{diag}({\cal A}_2,{\cal A}_3, \ldots, {\cal A}_{D-1})\ \quad \hbox{where} \quad
{\textstyle \sum_{i=2}^{D-1} {\cal A}_i=0} \,.
\end{equation}
In view of (\ref{Kundt geodesic deviationB_typeN}), the relative motion of (initially static) test particles is such that they \emph{recede} in spatial directions with \emph{positive eigenvalues} ${{\cal A}_i>0}$ , while they \emph{converge} with \emph{negative eigenvalues} ${{\cal A}_i<0}$ . In the directions where ${{\cal A}_i=0}$ the particles stay fixed.

\begin{figure}[t]
  \begin{center}
  \includegraphics[width=0.82\textwidth]{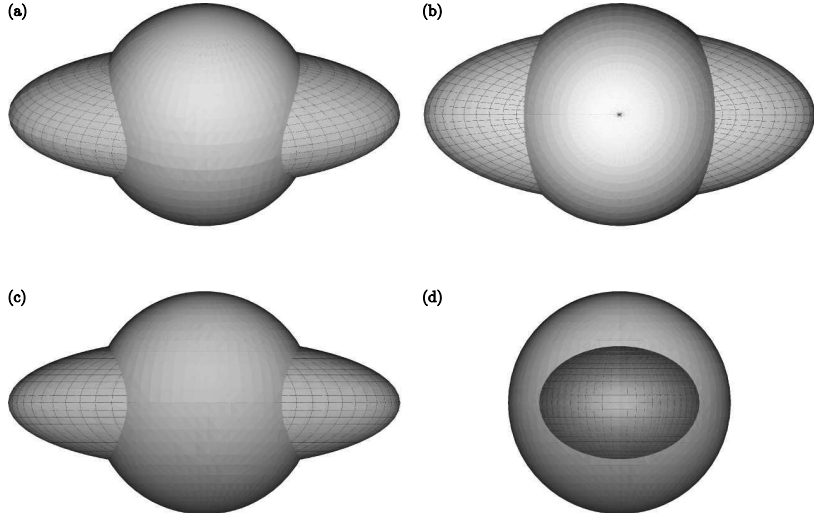}
  \end{center}
   \vspace{-3mm}
  \caption{\label{figure3} Deformation of a sphere of test particles in the case when one eigenvalue of ${-\Psi_{4^{ij}}^\nat}$ is positive and two are negative.}
\end{figure}

In the classic ${D=4}$ case, there is \emph{just one possibility}, namely ${{\cal A}_3=-{\cal A}_2}$, and the diagonalized matrix of the gravitational wave amplitudes takes the form
\begin{equation}
-\Psi_{4^{ij}}^\nat = \left(
\begin{array}{cc}
    {\cal A}_2 &  0 \\
    0 &  -{\cal A}_2 \\
  \end{array}
\right).
\end{equation}
In the transverse 2-dimensional space perpendicular to the propagation direction $\bolde_{(1)}$, we observe the standard gravitational wave effect, in which the set of test particles expands in the direction $\bolde_{(2)}$ when ${{\cal A}_2>0}$ and simultaneously contracts \emph{by the same amount} in the perpendicular direction $\bolde_{(3)}$ (or vice versa if ${{\cal A}_2<0}$), unless one has the trivial case ${{\cal A}_2=0}$.

\begin{figure}[t]
  \begin{center}
  \includegraphics[width=0.82\textwidth]{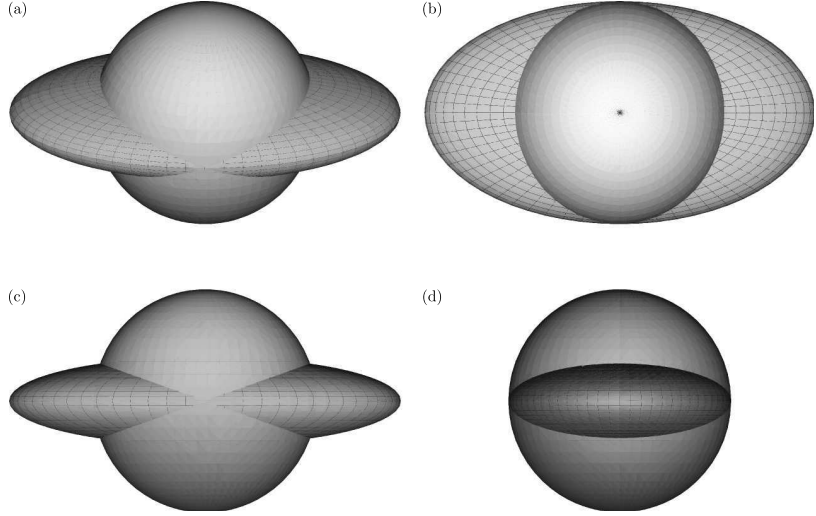}
  \end{center}
   \vspace{-3mm}
  \caption{\label{figure4} Deformation of a sphere of test particles in the case when one eigenvalue of ${-\Psi_{4^{ij}}^\nat}$
   is positive, one is zero and one is negative.}
\end{figure}

In higher dimensions, many more possibilities and new observable effects arise. For example, in the first non-trivial case ${D=5}$, the corresponding transverse space is 3-dimensional. Concerning the deformation of a 3-dimensional test sphere, there are \emph{three physically distinct situations} determined by ${{\cal A}_2,{\cal A}_3, {\cal A}_4}$, namely:

\begin{itemize}
\item two eigenvalues are positive and one is negative, see figure~\ref{figure2},

\item one eigenvalue is positive and two are negative, see figure~\ref{figure3},

\item one eigenvalue is positive, one is zero and one is negative, see figure~\ref{figure4}.
\end{itemize}

\noindent
From the point of view of a gravitational wave interferometric detector located in our (1+3)-dimensional ``real'' universe locally spanned by the vectors $(\boldu,\bolde_{(1)},\bolde_{(2)},\bolde_{(3)})$ where $\bolde_{(1)}$ is the propagation direction, $(\bolde_{(2)},\bolde_{(3)})$ defines the plane of the detector, while $\bolde_{(4)}$ is the extra (directly unobservable) dimension, we would see the following ``peculiar'' effects in which the usual \emph{traceless property in $(\bolde_{(2)},\bolde_{(3)})$ is violated}:

\begin{itemize}

\item  both ${{\cal A}_2, {\cal A}_3\not=0}$  (either positive or negative),  ${{\cal A}_4=-({\cal A}_2+{\cal A}_3)}$ due to (\ref{diagtrace}):
\begin{equation}
-\Psi_{4^{ij}}^\nat = \left(
\begin{array}{cc|c}
      {\cal A}_2     &  0  &   0 \\
    0 &   {\cal A}_3 &   0 \\
    \hline
    0 &   0 &  -({\cal A}_2 +{{\cal A}_3}) \\
  \end{array}
 \right).
\end{equation}
In the directly observable first sector of dimension ${2 \times 2}$, the eigenvalues ${\cal A}_2$ and~${\cal A}_3$ can have \emph{arbitrary} values now. Thus, test particles in these directions may 

\begin{itemize}
\item
recede in both directions $\bolde_{(2)},\bolde_{(3)}$ (${{\cal A}_2, {\cal A}_3>0\Rightarrow {\cal A}_4<0}$) as in figure~\ref{figure2}(b),

\item
converge in both directions $\bolde_{(2)},\bolde_{(3)}$ (${{\cal A}_2, {\cal A}_3<0\Rightarrow {\cal A}_4>0}$) as in figure~\ref{figure3}(d),

\item
recede in one direction and converge in the other, but not by the same amount (${\cal A}_2$, ${\cal A}_3$ have opposite signs, ${|{\cal A}_2|\not=|{\cal A}_3|\Rightarrow {\cal A}_4\not=0}$) as in figures~\ref{figure2}(c),~\ref{figure2}(d) or figures~\ref{figure3}(b),~\ref{figure3}(c),

\item behave as in the standard ${D=4}$ general relativity (${{\cal A}_3=-{\cal A}_2\Rightarrow {\cal A}_4=0}$) as in figure~\ref{figure4}(c), that is
\begin{equation}
-\Psi_{4^{ij}}^\nat = \left(
\begin{array}{cc|c}
    {\cal A}_2 &       0    &  0 \\
    0             & -{\cal A}_2 &  0 \\
    \hline
    0             &       0    &  0 \\
  \end{array}
\right).
\end{equation}

\end{itemize}

\item ${{\cal A}_3=0}$  or ${{\cal A}_2=0}$, so that

\begin{equation}
\hspace{-18mm}
-\Psi_{4^{ij}}^\nat = \left(
\begin{array}{cc|c}
    {\cal A}_2 &     0    &  0 \\
    0             &     0    &  0 \\
    \hline
    0             &     0  &  -{\cal A}_2 \\
  \end{array}
 \right)
  \quad \hbox{or}\quad
-\Psi_{4^{ij}}^\nat = \left(
\begin{array}{cc|c}
    0 &     0    &  0 \\
    0             &   {\cal A}_3   &  0 \\
    \hline
    0             &     0  &  -{\cal A}_3 \\
  \end{array}
\right).
\end{equation}
We can distinguish two subcases of this anomalous behaviour, namely
\begin{itemize}
\item
${{\cal A}_2>0}$ or ${{\cal A}_3>0}$ as in figure~\ref{figure4}(b),

\item
${{\cal A}_2<0}$ or ${{\cal A}_3<0}$ as in figure~\ref{figure4}(d).

\end{itemize}
\end{itemize}

\noindent
Finally, in principle, it may also happen that the gravitational wave would propagate in the \emph{extra} spatial dimension~$\bolde_{(4)}$, say. Due to the formal swap $\bolde_{(1)}\leftrightarrow\bolde_{(4)}$, this would imply ${\,{\cal A}_4=0\,}$ and ${\, {\cal A}_1\not=0\,}$: in our real universe we would thus observe a \emph{longitudinal deformation} of a cloud of test particles due to such higher-dimensional gravitational~wave.

\subsection{Type~III Kundt spacetimes}
\label{typIII}

For type~III Kundt spacetimes (\ref{obecny Kundt II}) (which have a triple WAND ${\mathbf{k}=\partial_r}$), all the Weyl tensor components of the boost weight ${0}$ vanish, ${\Psi_{2S}^\nat=\Psi_{2T^{ij}}^\nat=\Psi_{2^{ij}}^\nat=\Psi_{2^{ijkl}}^\nat=0}$.  The equations of geodesic deviation  (\ref{Kundt geodesic deviationA}), (\ref{Kundt geodesic deviationB}) thus become
\begin{eqnarray}
&&\hspace{-1mm}
\ddot{Z}^{(1)} = \frac{2\Lambda}{(D-2)(D-1)}\,Z^{(1)}
- \dot{u}\,\Psi_{3T^{j}}^\nat\,Z^{(j)} \, , \label{Kundt geodesic deviationA_typeIII}\\
&&\hspace{-0.6mm}
\ddot{Z}^{(i)} = \frac{2\Lambda}{(D-2)(D-1)}\,Z^{(i)}\,
- \dot{u}\,\Psi_{3T^{i}}^\nat\,Z^{(1)}  -\dot{u}^2\,\Psi_{4^{ij}}^\nat\,Z^{(j)}  \nonumber\\
&&\hspace{24.0mm}
 -2\,\dot{u}\,\dot{x}^p g_{pq}\big(\Psi_{3T^{(i}}^\nat m^q_{j^{)}}-\Psi_{3^{(ij)k}}^\nat m^{kq}\big) Z^{(j)}\, .\label{Kundt geodesic deviationB_typeIII}
\end{eqnarray}
Using the conditions summarized in the first four rows of table~\ref{classifKundt}, expressions (\ref{Psi 3TjD})--(\ref{Psi4D}) for the non-trivial Weyl scalars ${\Psi_{3T^j}^\nat}$, ${\Psi_{3^{ijk}}^\nat}$, ${\Psi_{4^{ij}}^\nat}$  reduce to
\begin{eqnarray}
&&\hspace{-20mm}
\Psi_{3T^j}^\nat = -m_{j}^{p}\,\frac{D-3}{D-2}\,\bigg[(a_{,p}+f_{p}\,a)\,r  \nonumber\\
&&\hspace{-8mm}   +\frac{1}{2}\bigg(b_{,p}-f_{p,u}-\frac{2\,e_{p}}{D-2}\Big(\frac{\,^{S}\!R}{D-3}+f\Big)+\frac{1}{2}f^qe_qf_p-f^q E_{qp}-\frac{2}{D-3}\,X_p\bigg)\bigg]
\, , \nonumber\\
&&\hspace{-20mm}
\Psi_{3^{ijk}}^\nat =\tilde\Psi_{3^{ijk}}^\nat+ \frac{1}{D-3}\,(\delta_{ij}\,\Psi_{3T^k}^\nat-\delta_{ik}\,\Psi_{3T^j}^\nat) \, , \\
&&\hspace{-20mm}
\tilde\Psi_{3^{ijk}}^\nat = m_{i}^{p}m_{j}^{m}m_{k}^{q}\,\Big(X_{pmq}-\frac{1}{D-3}\left( g_{pm}\,X_q-g_{pq}\,X_m\right)\Big) \, ,\nonumber\\
&&\hspace{-19mm}
\Psi_{4^{ij}}^\nat= m_{i}^{p}m_{j}^{q}\, \Big(W_{pq}-\frac{1}{D-2}\, g_{pq}\,W\Big)\,,\nonumber
\end{eqnarray}
where ${\,a=\frac{1}{4}f^p f_p-\frac{1}{D-2}\left(\frac{\,^{S}\!R}{D-3}+f\right)}$,
${\,X_q \equiv g^{pm}X_{pmq}\,}$,  ${\,W\equiv g^{pq}\,W_{pq}\,}$ and
\begin{eqnarray}
&&\hspace{-19mm}
X_{pmq} = {\textstyle e_{[q||m]||p}+e_{p[m}f_{q]}-f_{p[m}\,e_{q]}+g_{p[m,u||q]}}\, , \label{defXpmqspec}\\
&&\hspace{-17mm}
W_{pq} = {\textstyle
-r^2\left[\,\frac{1}{2}a_{||p||q}+\frac{1}{2}af_{(p||q)}+\frac{3}{2}a_{,(p}f_{q)}+a f_p f_q\,\right]} \nonumber\\
&&\hspace{-6mm} {\textstyle
-r\,\left[\,\frac{1}{2}b_{||p||q}+\frac{1}{2}b_{,(p}f_{q)}+ae_{pq}+2a f_{(p}e_{q)}+2a_{,(p}e_{q)}-
f_{(p}f_{q),u}-f_{(p,u||q)} \right]}
\nonumber\\
&&\hspace{-6mm} {\textstyle
-\frac{1}{2}c_{||p||q}+\frac{1}{2}cf_{(p||q)}+\frac{1}{2}c_{,(p}f_{q)}-\frac{1}{2}be_{pq} -b_{,(p}e_{q)}-ae_pe_q  }
\nonumber\\
&&\hspace{-6mm}
 {\textstyle
+e_{(p}f_{q),u}+e_{(p,u||q)}-\frac{1}{2}g_{pq,uu}+g^{mn}E_{mp}E_{nq}
+ f^m E_{m(p}\,e_{q)}-e^mE_{m(p}\,f_{q)}
 }\nonumber\\
&&\hspace{-6mm}
{\textstyle
+\frac{1}{4}(e^me_mf_pf_q+f^mf_me_pe_q)-\frac{1}{2}f^me_m f_{(p}e_{q)}  }
\, . \nonumber
\end{eqnarray}
In addition to the isotropic influence of the cosmological constant $\Lambda$ and the transverse effects of gravitational waves described by $\Psi_{4^{ij}}^\nat$ (which are typical for type~O and~N spacetimes, respectively), type~III Kundt spacetimes feature a \emph{longitudinal effect} proportional to the scalars $\Psi_{3T^{j}}^\nat$, see (\ref{Kundt geodesic deviationA_typeIII}). Moreover, from (\ref{Kundt geodesic deviationB_typeIII}) we conclude that there is also an additional \emph{kinematic effect for non-static observers} --- those with a non-vanishing velocity in the transverse space, ${\dot{x}^p\not=0}$. Measuring relative motions between geodesic observers with non-trivial spatial velocities we can thus determine other components of the curvature tensor, namely the symmetric part of ${\Psi_{3^{ijk}}^\nat}$.

If, and only if, ${\Psi_{4^{ij}}^\nat=0}$, the geometry is of algebraic type~III$_i$ with respect to the triple WAND $\boldk^\nat$ and WAND $\boldl^\nat$.

\subsection{Subtype~III(a)}
\label{subtypIIIa}
For the subtype III(a) of Kundt spacetimes, there is ${\Psi_{3T^j}^\nat =0}$, see  table~\ref{classifKundt} and~\cite{PodolskySvarc:2013}. The equations of geodesic deviation (\ref{Kundt geodesic deviationA_typeIII}), (\ref{Kundt geodesic deviationB_typeIII}) thus simplify to
\begin{eqnarray}
&&\hspace{-13mm}
\ddot{Z}^{(1)} = \frac{2\Lambda}{(D-2)(D-1)}\,Z^{(1)} \, , \label{Kundt geodesic deviationA_typeIIIa}\\
&&\hspace{-12.6mm}
\ddot{Z}^{(i)} = \frac{2\Lambda}{(D-2)(D-1)}\,Z^{(i)}\,
 -\dot{u}^2\,\Psi_{4^{ij}}^\nat\,Z^{(j)}
 +2\,\dot{u}\,\dot{x}^p g_{pq}\tilde\Psi_{3^{(ij)k}}^\nat m^{kq}\, Z^{(j)}\, .\label{Kundt geodesic deviationB_typeIIIa}
\end{eqnarray}
Apart from the cosmological background $\Lambda$-term, there are \emph{only transverse effects} given by the scalars $\Psi_{4^{ij}}^\nat$ and $\tilde\Psi_{3^{(ij)k}}^\nat$. The latter contribution is purely kinematical, i.e., it is absent for ${\dot x^p=0}$. For such static observers, the geodesic deviation is the same as for type~N spacetimes, cf. (\ref{Kundt geodesic deviationA_typeN}), (\ref{Kundt geodesic deviationB_typeN}). The specific contributions of $\tilde\Psi_{3^{(ij)k}}^\nat$ can be identified by considering non-static observers with mutual  velocities ${\dot x^p\not=0}$.

\subsection{Subtype~III(b)}
\label{subtypIIIb}
For the subtype III(b) of Kundt spacetimes, there is ${\tilde\Psi_{3^{ijk}}^\nat =0}$, see  table~\ref{classifKundt} and~\cite{PodolskySvarc:2013}.  In such a case, the equations (\ref{Kundt geodesic deviationA_typeIII}), (\ref{Kundt geodesic deviationB_typeIII}) become
\begin{eqnarray}
&&\hspace{-13mm}
\ddot{Z}^{(1)} = \frac{2\Lambda}{(D-2)(D-1)}\,Z^{(1)}
- \dot{u}\,\Psi_{3T^{j}}^\nat\,Z^{(j)} \, , \label{Kundt geodesic deviationA_typeIIIb}\\
&&\hspace{-12.6mm}
\ddot{Z}^{(i)} = \frac{2\Lambda}{(D-2)(D-1)}\,Z^{(i)}\,
- \dot{u}\,\Psi_{3T^{i}}^\nat\,Z^{(1)}  -\dot{u}^2\,\Psi_{4^{ij}}^\nat\,Z^{(j)}  \nonumber\\
&&\hspace{5.2mm}
 -2\,\frac{D-2}{D-3}\,\dot{u}\,\dot{x}^p g_{pq}\Big(\Psi_{3T^{(i}}^\nat m^q_{j^{)}}-\frac{\delta_{ij}}{D-2}\,\Psi_{3T^k}^\nat m^{kq}\Big) Z^{(j)}\, .\label{Kundt geodesic deviationB_typeIIIb}
\end{eqnarray}
The geodesic deviation is thus fully determined by the scalars $\Lambda$, $\Psi_{4^{ij}}^\nat$ and $\Psi_{3T^{j}}^\nat$ via the corresponding isotropic, transverse and longitudinal effects, respectively. For observers with non-vanishing spatial velocities ${\dot x^p\not=0}$, transverse motion is modified by the presence of $\Psi_{3T^{j}}^\nat$. This \emph{additional effect is traceless} since
${\delta^{ij}\,\Psi_{3T^{(i}}^\nat m^q_{j^{)}}=\Psi_{3T^i}^\nat m^{iq}}$.

\subsection{Type~D Kundt spacetimes}
\label{typD}

For type~D Kundt spacetimes (\ref{obecny Kundt II}) with a double WAND ${\boldk^\nat=\mathbf{k}=\partial_r}$  and a double WAND ${\boldl^\nat=\frac{1}{2}g_{uu}\,\mathbf{\partial}_r+\mathbf{\partial}_u}$, all the Weyl scalars $\Psi_{A^{...}}^\nat$ vanish, except for the boost weight~$0$. Therefore, the equations of geodesic deviation are (\ref{Kundt geodesic deviationA}), (\ref{Kundt geodesic deviationB}) with
\begin{eqnarray}
&&\hspace{-19.9mm}
\Psi_{2S} = \Psi_{2S}^\nat \, , \qquad \Psi_{2T^{(ij)}} = \Psi_{2T^{(ij)}}^\nat\,, \nonumber\\
&&\hspace{-21.2mm}
\Psi_{3T^i} =\sqrt{2}\,\dot{x}^p\,g_{pq}\,\Big(\big(\Psi_{2^{ij}}^\nat -\Psi_{2T^{ji}}^\nat\big) \,m^{jq}-\Psi_{2S}^\nat \,m_i^q\Big)\,, \nonumber\\
&&\hspace{-20.2mm}
\Psi_{4^{ij}} =2\,\dot{x}^p\dot{x}^q\Big(
\,g_{pq}\,\Psi_{2T^{(ij)}}^\nat -g_{pm}\,g_{qn}\,\Psi_{2S}^\nat\, m_i^m m_j^n +g_{pm}\,g_{qn}\,\Psi_{2^{ikjl}}^\nat m^{km} m^{ln}
 \label{PsinatnD} \\
&& \hspace{18mm}
-2\,g_{pm}\,g_{qn}\,\big(\Psi_{2^{k(i}}^\nat +\Psi_{2T^{k(i}}^\nat \big)\,m^n_{j^{)}} m^{km}\Big)
\, ,\nonumber
 \end{eqnarray}
where ${\Psi_{2S}^\nat}$, ${\Psi_{2T^{ij}}^\nat}$, ${\Psi_{2^{ij}}^\nat}$, ${\Psi_{2^{ijkl}}^\nat}$ are explicitly given by expressions (\ref{Psi2s})--(\ref{Psi2ij}).

For \emph{static observers} that do not move in the transverse spatial directions (${\dot x^p=0}$), we have ${\Psi_{3T^i}=0=\Psi_{4^{ij}} }$, so that the equations simplify considerably to
\begin{eqnarray}
&&\hspace{0mm}
\ddot{Z}^{(1)} = \frac{2\Lambda}{(D-2)(D-1)}\,Z^{(1)}+\Psi_{2S}^\nat\,Z^{(1)}\, , \label{Kundt geodesic deviationADstat}\\
&&\hspace{0mm}
\ddot{Z}^{(i)} = \frac{2\Lambda}{(D-2)(D-1)}\,Z^{(i)}-\Psi_{2T^{(ij)}}^\nat\,Z^{(j)}\, .\label{Kundt geodesic deviationBDstat}
\end{eqnarray}
We can now explicitly discuss specific particle motion in various algebraic subtypes of the Kundt spacetimes of type~D:

\subsection{Subtype~D(a)}
\label{subtypDa}
The subtype D(a) is defined by the condition
\begin{equation}
\Psi_{2S}^\nat=0\,. \label{condDa}
\end{equation}
This is equivalent to
${g_{uu}=a\,r^2+ b\,r+c\,}$ where ${a=\frac{1}{4}f^p f_p-\frac{1}{D-2}\big(\frac{\,^{S}\!R}{D-3}+f\big)}$,
cf. (\ref{Psi2s}) and  the first row in table~\ref{classifKundt}. The geodesic deviation equations (\ref{Kundt geodesic deviationADstat}), (\ref{Kundt geodesic deviationBDstat}) reduce to
\begin{eqnarray}
&&\hspace{0mm}
\ddot{Z}^{(1)} = \frac{2\Lambda}{(D-2)(D-1)}\,Z^{(1)} \, , \label{Kundt geodesic deviationADstatIIa}\\
&&\hspace{0mm}
\ddot{Z}^{(i)} = \frac{2\Lambda}{(D-2)(D-1)}\,Z^{(i)}-\Psi_{2T^{(ij)}}^\nat\,Z^{(j)}\, ,\label{Kundt geodesic deviationBDstatIIa}
\end{eqnarray}
where, in view of (\ref{Psi2T(ij)}), (\ref{condDa}), we have ${\Psi_{2T^{(ij)}}^\nat = \tilde\Psi_{2T^{(ij)}}^\nat}$ which is explicitly expressed by (\ref{tildePsi2T(ij)}). There is \emph{no longitudinal Newtonian motion}, see (\ref{Kundt geodesic deviationADstatIIa}), and the transverse New-tonian deformations (\ref{Kundt geodesic deviationBDstatIIa}) are traceless since
${\delta^{ij}\,\Psi_{2T^{(ij)}}^\nat =\Psi_{2T^{i}}^\nat{}^{_i}=\Psi_{2S}^\nat =0}$, see (\ref{constraints}). Interestingly, in higher dimensions, the local behaviour of test particles in subtype~D(a) spacetimes, as given by expressions (\ref{Kundt geodesic deviationADstatIIa}), (\ref{Kundt geodesic deviationBDstatIIa}), is very similar to the effect caused by type~N gravitational waves (\ref{Kundt geodesic deviationA_typeN}), (\ref{Kundt geodesic deviationB_typeN}). Due to this close formal similarity, we can use figures~\ref{figure2}--\ref{figure4} to illustrate particle motion in the ${D=5}$ case. Such a situation does \emph{not} appear in the ${D=4}$ case since ${\Psi_{2T^{(ij)}}^\nat = \tilde\Psi_{2T^{(ij)}}^\nat=0}$, as we can see from (\ref{tildePsi2T(ij)}).

For geodesics with spatial velocities ${\dot x^p\not=0}$, there are additional terms $\Psi_{3T^i}$, $\Psi_{4^{ij}}$ given by (\ref{Kundt geodesic deviationA}), (\ref{Kundt geodesic deviationB}), (\ref{PsinatnD}). The  scalars ${\Psi_{2^{ij}}^\nat}$, ${\Psi_{2^{ijkl}}^\nat}$ take the form (\ref{Psi2ij}) and (\ref{Psi2ijkl}), (\ref{tildePsi2ijkl}).

\subsection{Subtype~D(b)}
\label{subtypDb}
\begin{figure}[t]
  \begin{center}
  \includegraphics[width=0.82\textwidth]{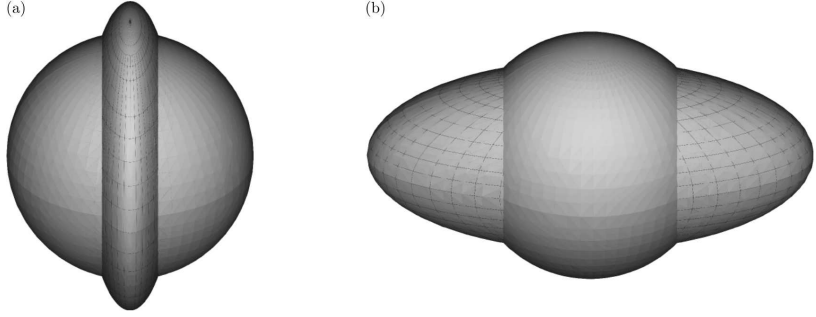}
  \end{center}
   \vspace{-3mm}
  \caption{\label{figure5} Deformation of a sphere of static test particles in the subtype~D(b) when ${D=5}$ for the cases (a) ${\Psi_{2S}^\nat<0}$, (b) ${\Psi_{2S}^\nat>0}$. Unlike in figures~\ref{figure2}--\ref{figure4}, here we show $\bolde_{(1)},\bolde_{(2)},\bolde_{(3)}$, where $\bolde_{(1)}$ is the longitudinal direction (oriented horizontally) while $\bolde_{(2)},\bolde_{(3)}$ (plotted perpendicularly) are two directions of the transverse 3-space (the third equivalent transverse direction $\bolde_{(4)}$ is suppressed).
}
\end{figure}

The subtype D(b) occurs if, and only if, ${\tilde\Psi_{2T^{(ij)}}^\nat=0}$. From (\ref{Psi2T(ij)}) it thus follows that
\begin{equation}
\Psi_{2T^{(ij)}}^\nat={\textstyle\frac{1}{D-2}\,\delta_{ij}\,\Psi_{2S}^\nat}\,. \label{condDb}
\end{equation}
Due to (\ref{tildePsi2T(ij)}), this is equivalent to ${\,^{S}\!R_{pq}-\frac{1}{D-2}\,g_{pq}\,^{S}\!R
=-\frac{1}{2}(D-4)\big(f_{pq}-\frac{1}{D-2}\,g_{pq}\,f\big)}$, see \cite{PodolskySvarc:2013} and the second row in table~\ref{classifKundt}. For such Kundt geometries, the equations of geodesic deviation  (\ref{Kundt geodesic deviationADstat}), (\ref{Kundt geodesic deviationBDstat}) take the form
\begin{eqnarray}
&&\hspace{0mm}
\ddot{Z}^{(1)} = \frac{2\Lambda}{(D-2)(D-1)}\,Z^{(1)}+\Psi_{2S}^\nat\,Z^{(1)}\, , \label{Kundt geodesic deviationADstatIIb}\\
&&\hspace{0mm}
\ddot{Z}^{(i)} = \frac{2\Lambda}{(D-2)(D-1)}\,Z^{(i)}-\frac{\delta_{ij}}{D-2}\,\Psi_{2S}^\nat\,Z^{(j)}\, .\label{Kundt geodesic deviationBDstatIIb}
\end{eqnarray}
We can see that the Newtonian part of the gravitational field is now fully determined by a \emph{single scalar} $\Psi_{2S}^\nat$ given by (\ref{Psi2s}). Moreover, motion in the transverse spatial directions ${i,j=2,\ldots,D-1}$ is \emph{isotropic} (its sum is fully offset to zero by the longitudinal motion,  ${\delta^{ij}\,\Psi_{2T^{(ij)}}^\nat =\Psi_{2S}^\nat}$). A sphere of test particles, initially at rest, is thus deformed into a \emph{rotational ellipsoid} with the axis $\bolde_{(1)}$, see figure~\ref{figure5}. Interestingly, this type of behaviour enables us to \emph{determine experimentally the dimension} $D$ of the spacetime. Subtracting the isotropic motion given by $\Lambda$, it is possible to measure the relative acceleration in the longitudinal direction $\bolde_{(1)}$ and compare it with the acceleration in \emph{any} transverse direction $\bolde_{(2)}$, say, obtaining ${(\ddot{Z}^{(1)}/Z^{(1)})/(-\ddot{Z}^{(2)}/Z^{(2)}) = D-2}$.

For geodesics with ${\dot x^p\not=0}$, the additional terms $\Psi_{3T^i}$ and $\Psi_{4^{ij}}$ given by (\ref{Kundt geodesic deviationA}), (\ref{Kundt geodesic deviationB}), (\ref{PsinatnD}) have to be included.

\subsection{Subtype~D(ab)}
\label{subtypDab}
Any Kundt spacetime that is \emph{both} of the algebraic subtype D(a) \emph{and} subtype D(b) must necessarily satisfy
${\Psi_{2S}^\nat=0= \Psi_{2T^{(ij)}}^\nat}$, see (\ref{condDa}) and (\ref{condDb}). Equations of geodesic deviation (\ref{Kundt geodesic deviationA}), (\ref{Kundt geodesic deviationB}) then reduce to
\begin{eqnarray}
&&\hspace{-15mm}
\ddot{Z}^{(1)} = \frac{2\Lambda}{(D-2)(D-1)}\,Z^{(1)}
- \frac{1}{\sqrt{2}}\,\Psi_{3T^j}\,Z^{(j)} \, , \label{Kundt geodesic deviationADab}\\
&&\hspace{-14.6mm}
\ddot{Z}^{(i)} = \frac{2\Lambda}{(D-2)(D-1)}\,Z^{(i)}
- \frac{1}{\sqrt{2}}\,\Psi_{3T^i}\,Z^{(1)} -\frac{1}{2}\,\Psi_{4^{ij}}\,Z^{(j)}\, ,\label{Kundt geodesic deviationBDab}
\end{eqnarray}
where (\ref{PsinatnD}) becomes
\begin{eqnarray}
&&\hspace{-15mm}
\Psi_{3T^i} =\frac{3}{2}\sqrt{2}\,m_{i}^{p}\,F_{pq}\,\dot{x}^q \,, \quad
\Psi_{4^{ij}} =2\, m_{i}^{m}m_{j}^{n}\,\big(\!  \,^{S}\!C_{mpnq}
-3F_{p(m}\,g_{n)q}\big)\,\dot{x}^p\dot{x}^q\, . \label{PsinatnDab}
 \end{eqnarray}
Interestingly, for \emph{static observers} (${\dot x^q=0}$), we have ${\Psi_{3T^i}=0=\Psi_{4^{ij}} }$ and the equations contain only the cosmological constant $\Lambda$-term. The relative motion of such test particles is \emph{the same as in the type~O spacetimes}
(\ref{Kundt geodesic deviationB_typeO}) --- it is fully isotropic as in the background Minkowski, de Sitter or anti-de Sitter spaces.

Recall also that in the classic ${D=4}$ case, the subtypes D(ab) and D(a) are identical because the condition for the subtype D(b) is always satisfied \cite{PodolskySvarc:2013}.

\subsection{Subtype~D(c)}
\label{subtypDc}
The algebraic subtype D(c) is defined by the condition ${\tilde\Psi^\nat_{2^{ikjl}}=0}$ which, using (\ref{tildePsi2ijkl}), is equivalent to ${\,^{S}\!C_{mpnq}=0 }$, cf. the third row in table~\ref{classifKundt}. Since ${\Psi_{2S}^\nat}$ and ${\Psi_{2T^{(ij)}}^\nat}$ are generally non-vanishing, this subtype of the Kundt geometries cannot be distinguished by measuring the deviation (\ref{Kundt geodesic deviationADstat}), (\ref{Kundt geodesic deviationBDstat}) of \emph{static} geodesic observers. In principle, it can be detected in the relative motion of non-static particles with ${\dot x^p\not=0}$ as the  $\tilde\Psi^\nat_{2^{ikjl}}$ component in the amplitude $\Psi_{4^{ij}}$ determined by (\ref{PsinatnD}) is \emph{absent}.

Moreover, as discussed in \cite{PodolskySvarc:2013}, the condition for subtype D(c) is identically satisfied in the cases ${D=4}$ and ${D=5}$.

\subsection{Subtype~D(d)}
\label{subtypDd}
The subtype D(d) occurs if, and only if, ${\Psi^\nat_{2^{ij}}=0}$. In view of (\ref{Psi2ij}), this is equivalent to ${F_{pq}=0}$ (see table~\ref{classifKundt}). As in the subcase D(c), this is not directly observable in the geodesic deviation (\ref{Kundt geodesic deviationADstat}), (\ref{Kundt geodesic deviationBDstat}) of static observers, but it is implied by the absence of the $\Psi^\nat_{2^{ij}}$ component entering the scalars $\Psi_{3T^i}$, $\Psi_{4^{ij}}$ via (\ref{PsinatnD}). It is detectable by observers with ${\dot x^p\not=0}$ for which the equations of geodesic deviation take the form (\ref{Kundt geodesic deviationA}), (\ref{Kundt geodesic deviationB}).

\subsection{Type~II Kundt spacetimes}
\label{typII}

The general form of the geodesic deviation equations for any Kundt spacetime  (\ref{obecny Kundt II}) of algebraic type~II (or more special) with (at least) a double WAND ${\mathbf{k}=\boldk^\nat=\mathbf{\partial}_r}$ is
\begin{eqnarray}
&&\hspace{-25mm}
\ddot{Z}^{(1)} = \frac{2\Lambda}{(D-2)(D-1)}\,Z^{(1)} +\Psi^\nat_{2S}\,Z^{(1)} \nonumber \\
&&\hspace{-15mm}
-\Big[\dot{u}\Psi^\nat_{3T^j} +\dot{x}^pg_{pq}\Big(\frac{3}{2}\Psi^\nat_{2^{ji}}m^{iq}
-\tilde\Psi^\nat_{2T^{(ij)}}m^{iq}-\frac{D-1}{D-2}\Psi^\nat_{2S}m_j^q\Big)\Big]Z^{(j)}\,, \label{geoIIa}\\
&&\hspace{-25mm}
\ddot{Z}^{(i)} = \frac{2\Lambda}{(D-2)(D-1)}\,Z^{(i)}
-\Big(\tilde\Psi^\nat_{2T^{(ij)}}+\frac{\delta_{ij}}{D-2}\Psi^\nat_{2S}\Big)Z^{(j)}
-\dot{u}^2\Psi^\nat_{4^{ij}}\,Z^{(j)}  \nonumber \\
&&\hspace{-15mm}
-\Big[\dot{u}\Psi^\nat_{3T^i} +\dot{x}^pg_{pq}\Big(\frac{3}{2}\Psi^\nat_{2^{ij}}m^{jq}-\tilde\Psi^\nat_{2T^{(ij)}}m^{jq}
-\frac{D-1}{D-2}\Psi^\nat_{2S}m_i^q\Big)\Big]\,Z^{(1)}
\nonumber \\
&&\hspace{-15mm} -2\dot{u}\dot{x}^pg_{pq}\Big(\frac{D-2}{D-3}\Psi^\nat_{3T^{(i}}m_{j)}^q
-\frac{\delta_{ij}}{D-3}\Psi^\nat_{3T^{k}}m^{kq}-\tilde\Psi^\nat_{3^{(ij)k}}m^{kq}\Big)Z^{(j)} \nonumber \\
&&\hspace{-15mm}-\dot{x}^m\dot{x}^n\Big[
g_{mp}g_{nq}m^{kp}\big(\tilde\Psi^\nat_{2^{ikjl}}m^{lq}-3\Psi^\nat_{2^{k(i}} m_{j)}^q \big)
+\frac{D-2}{D-4}g_{mn}\tilde\Psi^\nat_{2T^{(ij)}} \nonumber \\
&&\hspace{-1mm}
+\frac{g_{mp}g_{nq}}{D-4}m^{kp}\Big(2\delta_{ij}\tilde\Psi^\nat_{2T^{(kl)}}m^{lq}
-(D-2)\big(\tilde\Psi^\nat_{2T^{(ik)}}m_j^q+\tilde\Psi^\nat_{2T^{(jk)}}m_i^q\big)\Big)\nonumber \\
&&\hspace{-1mm}
-\frac{D-1}{D-3}\Psi^\nat_{2S}\Big(g_{mp}g_{nq}m_i^pm_j^q-\frac{\delta_{ij}}{D-2}g_{mn}\Big)\Big]\,Z^{(j)} \,.\label{geoIIb}
\end{eqnarray}
The behavior of test particles in the subtypes II(a), II(b), II(c) and II(d) is easily obtained by setting ${\Psi_{2S}^\nat=0}$, ${\tilde\Psi_{2T^{(ij)}}^\nat=0}$, ${\tilde\Psi^\nat_{2^{ikjl}}=0}$ and ${\Psi^\nat_{2^{ij}}=0}$, respectively.

When all these Weyl scalars of the boost weight 0 vanish, we obtain the type~III Kundt geometries with a triple WAND  ${\mathbf{k}}$ and recover the results of sections~\ref{typIII}--\ref{subtypIIIb}. If, in addition, ${\Psi_{3T^j}^\nat=0=\tilde\Psi_{3^{ijk}}^\nat}$, the spacetimes are of type~N with a quadruple WAND  ${\mathbf{k}}$ discussed in~\ref{typN}, and with ${\Psi_{4^{ij}}^\nat=0}$ they become type~O, see~\ref{typO}. Alternatively, if only ${\Psi_{4^{ij}}^\nat=0}$, given by (\ref{Psi4D}), (\ref{WpqD}), (\ref{W}), the spacetime is of algebraic type~II$_i$ with respect to a double WAND  ${\boldk^\nat=\mathbf{\partial}_r}$ and WAND  ${\boldl^\nat=\frac{1}{2}g_{uu}\,\mathbf{\partial}_r+\mathbf{\partial}_u}$. When only the scalars ${\Psi_{3T^j}^\nat, \tilde\Psi_{3^{ijk}}^\nat}$ are non-trivial, the geometry is of algebraic type~III$_i$ with respect to a triple WAND $\boldk^\nat$  and WAND  $\boldl^\nat$.

Type~D Kundt geometries of section~\ref{typD} arise by setting ${\Psi^\nat_{3T^j}=0=\tilde\Psi^\nat_{3^{ijk}}}$ and ${\Psi^\nat_{4^{ij}}=0}$, in which case the expressions (\ref{geoIIa}), (\ref{geoIIb}) correspond to (\ref{PsinatnD}). The subtypes D(a), D(b), D(c) and D(d) are obtained when ${\Psi_{2S}^\nat=0}$, ${\tilde\Psi_{2T^{(ij)}}^\nat=0}$, ${\tilde\Psi^\nat_{2^{ikjl}}=0}$ and ${\Psi^\nat_{2^{ij}}=0}$, respectively, reducing the results to those discussed in sections~\ref{subtypDa}--\ref{subtypDd}.

\section{Example: type~II and~N gravitational waves on~D and~O backgrounds}
\label{example}

As an interesting illustration, we can consider a line element of the form
\begin{equation}
\dd s^2 = g_{pq} \,\dd x^p\dd x^q-2\,\dd u\,\dd r+(a\,r^2+c)\,\dd u^2 \, , \label{wave on Bert}
\end{equation}
where ${g_{pq}=g_{pq}(x)}$, ${a=\ }$const. and ${c=c(u,x)}$.

\goodbreak

\noindent
The possible algebraic structure of such Kundt geometries is summarized in table~\ref{classif wave on Bert}.
\begin{table}[ht]

\begin{tabular}{cc}
\hline\hline
Type & Necessary and sufficient conditions \\
\hline
II(a)  & ${\,a=-\frac{1}{(D-2)(D-3)}\,^{S}\!R}$ \\
II(b)  & ${\,^{S}\!R_{pq}=\frac{1}{D-2}\,g_{pq}\,^{S}\!R}$ \\
II(c)  & ${\,^{S}\!C_{mpnq}=0}$ \\
II(d)  & Always \\
\hline
N & II(abcd) \\
\hline
O & N with ${c_{||p||q}=\frac{1}{D-2}\,g_{pq}\,\triangle c}$ \\
\hline
D & ${c_{||p||q}=\frac{1}{D-2}\,g_{pq}\,\triangle c}$ \\
D(a) & D with II(a) \\
D(b) & D with II(b) \\
D(c) & D with II(c) \\
D(d) & D with II(d) \\
\hline\hline
\end{tabular}
\caption{\label{classif wave on Bert} The structure of all Kundt geometries (\ref{wave on Bert}) with respect to a multiple WAND ${\boldk^\nat=\partial_r}$ and (possibly double) WAND ${\boldl^\nat=\frac{1}{2}(a\,r^2+c)\mathbf{\partial}_r+\mathbf{\partial}_u}$.}
\end{table}

Relative motion of free test particles in these spacetimes is described by equations (\ref{Kundt geodesic deviationA}), (\ref{Kundt geodesic deviationB}) where the scalars (\ref{Psinatn}) take the form
\begin{eqnarray}
&&\hspace{-20.7mm}
 \Psi_{2S}= \frac{D-3}{D-1}\,\Big(\,a+\frac{\,^{S}\!R}{(D-2)(D-3)}\Big), \nonumber\\
&&\hspace{-20.7mm} \Psi_{2T^{(ij)}}=\frac{m_i^pm_j^q}{D-2}\,\Big(\,^{S}\!R_{pq}-\frac{g_{pq}}{D-2}\,^{S}\!R\,\Big)
+\frac{\delta_{ij}}{D-2}\frac{D-3}{D-1}\Big(\,a+\frac{\,^{S}\!R}{(D-2)(D-3)}\Big), \nonumber\\
&&\hspace{-20.7mm}
 \Psi_{3T^j}=-\frac{\sqrt{2}}{D-2}\,\dot{x}^pm_j^q\,\Big(\,^{S}\!R_{pq}+(D-3)\,a\,g_{pq}\,\Big), \\
&&\hspace{-20.7mm}
 \Psi_{4^{ij}}= -\dot{u}^2\,m_i^pm_j^q\,\Big(c_{||p||q}-\frac{g_{pq}}{D-2}\triangle c\Big) \nonumber\\
&&\hspace{-10.0mm}
+2\,\dot{x}^p\dot{x}^qm_i^mm_j^n\,\Big\{\,^{S}\!C_{mpnq}+
\,\Big(\,a+\frac{\,^{S}\!R}{(D-2)(D-3)}\Big)\Big(\frac{g_{pq}\,g_{mn}}{D-2}-g_{pm}g_{qn}\Big) \nonumber \\
&&\hspace{-6.0mm}
+\frac{1}{D-4}\,\Big[g_{pq}\Big(\,^{S}\!R_{mn}-\frac{g_{mn}}{D-2}\,^{S}\!R \Big)+\frac{2g_{mn}}{D-2}\Big(\,^{S}\!R_{pq}-\frac{g_{pq}}{D-2}\,^{S}\!R \Big) \nonumber\\
&&\hspace{5.5mm} -g_{pm}\Big(\,^{S}\!R_{qn}-\frac{g_{qn}}{D-2}\,^{S}\!R\Big)-g_{qn}\Big(\,^{S}\!R_{pm}
-\frac{g_{pm}}{D-2}\,^{S}\!R \Big)\Big]\Big\} \,. \nonumber\label{psiexample}
\end{eqnarray}
Notice that for the subtype~II(ab)$\equiv$II(abd), this simplifies considerably to
\begin{eqnarray}
&&\hspace{-20.2mm}
\ddot{Z}^{(1)} = \frac{2\Lambda}{(D-2)(D-1)}\,Z^{(1)} \,, \nonumber \\
&&\hspace{-20.2mm}
\ddot{Z}^{(i)} = \frac{2\Lambda}{(D-2)(D-1)}\,Z^{(i)} +\frac{1}{2}\,\dot{u}^2\,m_i^pm_j^q\,
 \Big(c_{||p||q}-\frac{g_{pq}}{D-2}\triangle c\Big)\,Z^{(j)}  \\
&&\hspace{21.5mm} -\dot{x}^p\dot{x}^q\,m_i^mm_j^n\,\,^{S}\!C_{mpnq}\,Z^{(j)}\,.\nonumber
\end{eqnarray}
When, in addition, ${\,^{S}\!C_{mpnq}=0}$, this becomes type~II(abcd)$\equiv$N.

If, and only if, ${c_{||p||q}=\frac{1}{D-2}\,g_{pq}\,\triangle c}$, the spacetimes are of type~D or type~O. When ${c=0}$, these belong to the important family of \emph{direct-product spacetimes}, see section~11 of \cite{PodolskySvarc:2013}, for which the first term of the metric (\ref{wave on Bert}) is a ${(D-2)}$-dimensional Riemannian space with metric $g_{pq}(x)$, while the second part is a 2-dimensional Lorentzian spacetime of constant Gaussian curvature~$a$. In general, $g_{pq}(x)$ need not be of constant curvature, but for the subtype D(a), $a$ is uniquely related to the \emph{constant} Ricci scalar ${\,^{S}\!R}$ of the transverse $(D-2)$-dimensional space. Such metrics represent natural higher-dimensional generalizations of the (anti-)Nariai, Pleba\'{n}ski--Hacyan, Bertotti--Robinson and Minkowski spacetimes of types~D or~O, see~\cite{GriffithsPodolsky:2009}.

For a non-trivial $c$, the spacetimes (\ref{wave on Bert}) are of type~II or of type~N. These can be naturally interpreted as the class of \emph{exact Kundt gravitational waves} with the profile $c(u,x)$ propagating in various direct-product background universes of algebraic types~D or~O mentioned above (and listed in table~6 of \cite{PodolskySvarc:2013}; see also \cite{PodolskyOrtaggio:2003,KadlecovaZelnikovKrtousPodolsky:2009,KrtousPodolskyZelnikovKadlecova:2012}).

The class of metrics (\ref{wave on Bert}) clearly contains \emph{pp-waves} (without gyratonic sources) propagating in flat space when ${a=0}$. These are of type~N if, and only if, ${g_{pq}=\delta_{pq}}$ (in which case they belong to the class of VSI spacetimes, see \cite{PodolskySvarc:2013}).

Finally, let us observe that in the classic ${D=4}$ case, the scalars (\ref{psiexample}) read
\begin{eqnarray}
&&\hspace{-25.2mm}
 \Psi_{2S}= \frac{1}{3}\,\Big(a+\frac{1}{2}\,^{S}\!R\,\Big)\,, \qquad \Psi_{2T^{(ij)}}=\frac{1}{6}\delta_{ij}\Big(a+\frac{1}{2}\,^{S}\!R\,\Big) \,, \nonumber\\
&&\hspace{-25.2mm}
 \Psi_{3T^j}=-\frac{\sqrt{2}}{2}\,\dot{x}^pm_j^q\,\big(\,^{S}\!R_{pq}+a\,g_{pq}\,\big) \,, \\
&&\hspace{-25.2mm}
 \Psi_{4^{ij}}= -\dot{u}^2\,m_i^pm_j^q \Big(c_{||p||q}-\frac{1}{2} g_{pq}\triangle c\Big)
+\dot{x}^p\dot{x}^qm_i^mm_j^n\Big(a+\frac{1}{2}\,^{S}\!R\,\Big)\Big(g_{pq}\,g_{mn}-2g_{pm}g_{qn}\Big) .\nonumber
\end{eqnarray}
The corresponding Kundt geometries (\ref{wave on Bert}) are thus generally of type~II$\equiv$II(bcd). They are of type~N$\equiv$II(abcd) if, and only if, ${\,a=-\frac{1}{2}\!\,^{S}\!R}$ with the only non-vanishing Weyl scalar ${ \Psi_{4^{ij}}= -\dot{u}^2\,m_i^pm_j^q \,(c_{||p||q}-\frac{1}{2} g_{pq}\triangle c)}$. In fact, this is the subfamily ${\alpha=\beta, \varepsilon=1, C=0}$ of spacetimes discussed in \cite{PodolskyOrtaggio:2003} and in sections 18.6--18.7 of \cite{GriffithsPodolsky:2009} (with the identification ${\alpha=\beta=\frac{1}{4}\!\,^{S}\!R}$, ${D=a}$ and ${H=-c}$) which was interpreted as exact Kundt gravitational waves of type II propagating on type~D backgrounds, and type N waves propagating on conformally flat type~O backgrounds, respectively. These background universes with the geometry of a direct product of two constant-curvature 2-spaces involve the standard Minkowski, Bertotti--Robinson, (anti-)Nariai and Pleba\'{n}ski--Hacyan spacetimes, cf. \cite{Ortaggio:2002,OrtaggioPodolsky:2002,KadlecovaZelnikovKrtousPodolsky:2009}.

\section{Conclusions}
\label{conclusions}

We systematically analyzed the general class of Kundt geometries in an arbitrary dimension ${D\ge4}$ using the geodesic deviation in Einstein's theory. We explicitly determined the specific motion of free test particles for all possible algebraically special spacetimes, including the corresponding subtypes, and demonstrated that the invariant quantities determining these (sub)types are measurable by detectors via characteristic relative accelerations. For example, the dimension of the spacetime can be measured directly by Newtonian-type tidal deformations of the algebraic subtype~D(b). The purely transverse type~N effects represent exact gravitational waves with ${D(D-3)/2}$ polarizations, which exhibit new and peculiar observable effects in higher dimensions ${D>4}$. We gave an example of such geometric and physical interpretation of the Kundt family by analyzing the class of type~N or~II gravitational waves propagating on backgrounds of type~O or~D.

\section*{Acknowledgements}

This work was supported by the grant GA\v{C}R P203/12/0118.


\section*{References}
\begin{thebibliography}{10}

\bibitem{Kundt:1961}
Kundt W 1961 The plane-fronted gravitational waves
{\em Z.~Physik} {\bf 163} 77--86

\bibitem{Kundt:1962}
Kundt W 1962 Exact solutions of the field equations: twist-free pure radiation
fields {\em Proc.~Roy.~Soc.~A} {\bf 270} 328--34

\bibitem{Stephani:2003}
Stephani~H, Kramer~D, MacCallum~M, Hoenselaers~C and Herlt~E 2003 {\em Exact Solutions of Einstein's Field
  Equations} (Cambridge: Cambridge University Press)

\bibitem{GriffithsPodolsky:2009}
Griffiths~J and Podolsk\'{y}~J 2009 {\em Exact Space-Times in Einstein's
  General Relativity} (Cambridge: Cambridge University Press)

\bibitem{Coley:2008}
Coley~A 2008 Classification of the Weyl tensor in higher dimensions and applications {\em Class. Quantum Grav.} {\bf 25} 033001 (29pp)

\bibitem{OrtaggioPravdaPravdova:2013}
Ortaggio~M, Pravda~V and Pravdov\'a~A 2013 Algebraic classification of higher dimensional spacetimes based on null alignment {\em Class. Quantum Grav.} {\bf 30} 013001 (57pp)

\bibitem{PodolskyZofka:2009}
Podolsk\'{y} J and \v{Z}ofka M 2009 General Kundt spacetimes in higher
  dimensions {\em Class. Quantum Grav.} {\bf 26} 105008 (18pp)

\bibitem{ColeyEtal:2009}
Coley A, Hervik S, Papadopoulos G and Pelavas N 2009 Kundt spacetimes
{\em Class. Quantum Grav.} {\bf 26} 105016 (34pp)

\bibitem{PodolskySvarc:2013}
Podolsk\'{y} J and \v{S}varc R 2013 Explicit algebraic classification of Kundt geometries in any dimension {\em Class. Quantum Grav.} {\bf 30} 125007 (25pp)

\bibitem{Bri25}
Brinkmann~H~W 1925 Einstein spaces which are mapped conformally on each other {\em Math. Annal.} {\bf 94} 119--45

\bibitem{ColMilPelPraPraZal03}
Coley~A, Milson~R, Pelavas~N, Pravda~V, Pravdov\'a~A and Zalaletdinov 2003 Generalizations of pp-wave spacetimes in higher dimensions {\em Phys. Rev.}~D {\bf 67} 104020

\bibitem{ColeyMilsonPravdaPravdova:2004}
Coley~A, Milson R, Pravda V and Pravdov\'a A 2004 Classification of the Weyl tensor in higher dimensions {\em Class. Quantum Grav.} {\bf 21} L35--41

\bibitem{ColMilPraPra04}
Coley~A, Milson~R, Pravda~V and Pravdov\'a~A 2004 Vanishing scalar invariant spacetimes in higher dimensions {\em Class. Quantum Grav.} {\bf 21} 5519--42

\bibitem{ColFusHerPel06}
Coley~A, Fuster~A, Hervik~S and Pelavas~N 2006 Higher dimensional VSI spacetimes {\em Class. Quantum Grav.} {\bf 23} 7431--44

\bibitem{ColHerPel06}
Coley~A, Hervik~S and Pelavas~N 2006 On spacetimes with constant scalar invariants {\em Class. Quantum Grav.} {\bf 23} 3053--74

\bibitem{ColHerPel09}
Coley~A, Hervik~S and Pelavas~N 2009 Spacetimes characterized by their scalar curvature invariants {\em Class. Quantum Grav.} {\bf 26} 025013 (33pp)

\bibitem{Bon70}
Bonnor~W~B 1970 Spinning null fluid in general relativity {\em Int.~J.~Theoret. Phys.} {\bf 3} 257--66

\bibitem{FroFur05}
Frolov~V~P and Fursaev~D~V 2005 Gravitational field of a spinning radiation beam pulse in higher dimensions {\em Phys. Rev.}~D {\bf 71} 104034

\bibitem{FroIsrZel05}
Frolov~V~P, Israel~W and Zelnikov~A 2005 Gravitational field of relativistic gyratons {\em Phys. Rev.}~D {\bf 72} 084031

\bibitem{FroZel05}
Frolov~V~P and Zelnikov~A 2005 Relativistic gyratons in asymptotically AdS spacetime {\em Phys. Rev.}~D {\bf 72} 104005

\bibitem{FroZel06}
Frolov~V~P and Zelnikov~A 2006 Gravitational field of charged gyratons {\em Class. Quantum Grav.} {\bf 23} 2119--28


\bibitem{CalLeZor07}
Caldarelli~M~M, Klemm~D and Zorzan~E 2007 Supersymmetric gyratons in five dimensions {\em Class. Quantum Grav.} {\bf 24} 1341--57

\bibitem{KadlecovaZelnikovKrtousPodolsky:2009}
Kadlecov\'{a}~H, Zelnikov~A, Krtou\v{s}~P and Podolsk\'{y}~J 2009 Gyratons on direct-product spacetimes {\em Phys. Rev. D} {\bf 80} 024004

\bibitem{KrtousPodolskyZelnikovKadlecova:2012}
Krtou\v{s}~P, Podolsk\'{y}~J, Zelnikov~A and Kadlecov\'{a}~H 2012 Higher-dimensional Kundt waves and gyratons {\em Phys. Rev. D} {\bf 86} 044039

\bibitem{NuttMilsonColey:2013}
McNutt~D, Milson~R and Coley~A 2013 Vacuum Kundt waves {\em Class. Quantum Grav.} {\bf 30} 055010 (29pp)

\bibitem{Wils:1989}
Wils~P 1989 Homogeneous and conformally Ricci flat pure radiation fields {\em Class.~Quantum Grav.} {\bf 6} 1243--51

\bibitem{KoutrasMcIntosh:1996}
Koutras~A and McIntosh~C 1996 A metric with no symmetries or invariants {\em Class.~Quantum Grav.} {\bf 13} L47--9

\bibitem{EdgarLudwig:1997a}
Edgar~S~B and Ludwig~G 1997 All conformally flat pure radiation metrics {\it Class.~Quantum Grav.} {\bf 14} L65--8

\bibitem{Skea:1997}
Skea~J~E~F 1997 The invariant classification of conformally flat pure radiation spacetimes {\it Class.~Quantum Grav.} {\bf 14} 2393--404

\bibitem{GriffithsPodolsky:1998}
Griffiths~J~B and Podolsk\'{y}~J 1998 Interpreting a conformally flat pure radiation space-time {\it Class.~Quantum Grav.} {\bf 15} 3863--71

\bibitem{Barnes:2001}
Barnes~A 2001 On the symmetries of the Edgar--Ludwig metric {\it Class.~Quantum Grav.} {\bf 18} 5287--91

\bibitem{WilsVandenBergh:1990}
Wils~P and Van den Bergh~N 1990 Petrov type D pure radiation fields of Kundt's class, {\em Class.~Quantum Grav.} {\bf 7} 577--80

\bibitem{GrooteBerghWylleman:2010}
De~Groote~L, Van den Bergh~N and Wylleman~L 2010 Petrov type~D pure radiation fields of Kundt's class {\em J.~Math.~Phys.} {\bf 51} 102501

\bibitem{Kinnersley:1969a}
Kinnersley~W 1969 Type~D vacuum metrics {\em J.~Math.~Phys.} {\bf 10} 1195--203

\bibitem{PlebanskiDemianski:1976}
 Pleba\'nski~J~F and Demia\'nski~M 1976 Rotating charged and uniformly accelerating mass in general relativity {\em Ann. Phys. (NY)} {\bf 98} 98--127

\bibitem{GriffithsPodolsky:2006b}
 Griffiths~J~B and Podolsk\'{y}~J 2006  A new look at the Pleba\'nski--Demia\'nski family of solutions {\em Int. J. Mod. Phys. D} {\bf 15} 335--69

\bibitem{OzsvathRobinsonRozga:1985}
Ozsv\'{a}th~I, Robinson~I and R\'{o}zga~K 1985 Plane-fronted gravitational and electromagnetic waves in spaces with cosmological constant {\em J.~Math.~Phys.} {\bf 26} 1755--61

\bibitem{Siklos:1985}
Siklos~S~T~C 1985 Lobatchevski plane gravitational waves, in {\em Galaxies, Axisymmetric Systems and Relativity} ed M~A~H~MacCallum (Cambridge: Cambridge University Press) 247--74

\bibitem{Podolsky:1998a}
Podolsk\'{y}~J 1998 Interpretation of the Siklos solutions as exact gravitational waves in the anti-de~Sitter universe {\em Class.~Quantum Grav.} {\bf 15} 719--33

\bibitem{BicakPodolsky:1999a}
Bi\v{c}\'{a}k~J and Podolsk\'{y} J 1999 Gravitational waves in vacuum spacetimes with cosmological constant. I. Classification and
geometrical properties of non-twisting type N solutions {\em J.~Math.~Phys.} {\bf 40} 4495--505

\bibitem{BicakPodolsky:1999b}
Bi\v{c}\'{a}k~J and Podolsk\'{y} J 1999 Gravitational waves in vacuum spacetimes with cosmological constant. II. Deviation of
geodesics and interpretation of non-twisting type N solutions {\em J.~Math.~Phys.} {\bf 40} 4506--17

\bibitem{GriffithsDochertyPodolsky:2004}
Griffiths~J~B, Docherty~P and Podolsk\'{y}~J 2004 Generalized Kundt waves and their physical interpretation {\em Class.~Quantum Grav.} {\bf 21} 207--22

\bibitem{PodolskyOrtaggio:2003}
Podolsk\'{y}~J and Ortaggio~M 2003 Explicit Kundt type II and N solutions as gravitational waves in various type D and O universes {\em Class.~Quantum Grav.} {\bf 20} 1685--701

\bibitem{PodolskyBelan:2004}
Podolsk\'{y}~J and Bel\'{a}\v{n}~M 2004 Geodesic motion in Kundt spacetimes and the character of the envelope singularity {\em Class.~Quantum Grav.} {\bf 21} 2811--29

\bibitem{LeviCivita:1926}
Levi-Civita~T 1926 Sur l'\'{e}cart g\'{e}od\'{e}sique
{\em Math. Ann.} {\bf 97} 291--320

\bibitem{Synge:1934}
Synge~J~L 1934 On the deviation of geodesics and null-geodesics, particularly in relation to the properties of spaces of constant curvature and indefinite line-element {\em Ann. Math.} {\bf 35} 705--13; reprinted in {\em Gen. Rel. Grav.} {\bf 41} (2009) 1205--14

\bibitem{Pirani:1956}
Pirani~F~A~E 1956 On the physical significance of the Riemann tensor
 {\em Acta Phys. Polon.} {\bf 15} 389--405; reprinted in {\em Gen. Rel. Grav.} {\bf 41} (2009) 1215--32

\bibitem{Szekeres:1965}
Szekeres~P 1965 The gravitational compass
 {\em J.~Math. Phys.} {\bf 6} 1387--91

\bibitem{FeliceBini:book}
de Felice~F and Bini~D 2010
{\em Classical Measurements in Curved Space-Times\/}
(Cambridge: Cambridge University Press)

\bibitem{PodolskySvarc:2012}
Podolsk\'{y} J and \v{S}varc R 2012 Interpreting spacetimes of any dimension
  using geodesic deviation {\em Phys. Rev. D} {\bf 85} 044057

\bibitem{KrtousPodolsky:2006}
Krtou\v{s} P and Podolsk\'{y} J 2006 Asymptotic structure of radiation in
  higher dimensions {\em Class. Quantum Grav.} {\bf 23} 1603--15

\bibitem{PravdaPravdovaColeyMilson:2004}
Pravda~V, Pravdov\'{a}~A, Coley~A and Milson~R 2004 Bianchi identities in
  higher dimensions {\em Class. Quantum Grav.} {\bf 21} 2873--98

\bibitem{PraPraOrt07}
Pravda~V, Pravdov\'a~A and Ortaggio~M 2007 Type D Einstein spacetimes in higher dimensions {\em Class. Quantum Grav.} {\bf 24} 4407--28

\bibitem{DurPraPraReall10}
Durkee~M, Pravda~V, Pravdov\'a~A and Reall~H~S 2010 Generalization of the Geroch--Held--Penrose formalism to higher dimensions
{\em Class. Quantum Grav.} {\bf 27} 215010 (21pp)

\bibitem{OrtaggioPravdaPravdova:2007}
Ortaggio~M, Pravda~V and Pravdov\'a~A 2007 Ricci identities in higher dimensions {\em Class. Quantum Grav.} {\bf 24} 1657--64

\bibitem{Ortaggio:2009}
Ortaggio~M 2009 Bel--Debever criteria for the classification of the Weyl tensor in higher dimensions {\em Class. Quantum Grav.} {\bf 26} 195015 (8pp)

\bibitem{ColeyHervik:2010}
Coley~A and Hervik~S 2010 Higher dimensional bivectors and classification of the Weyl operator {\em Class. Quantum Grav.} {\bf 27} 015002 (21pp)

\bibitem{Ortaggio:2002}
Ortaggio~M 2002 Impulsive waves in the Nariai universe {\em Phys.~Rev.~D} {\bf 65} 084046

\bibitem{OrtaggioPodolsky:2002}
Podolsk\'{y}~J and Ortaggio~M 2003 Impulsive waves in electrovac direct product spacetimes with~$\Lambda$ {\em Class.~Quantum Grav.} {\bf 19} 5221--7

\end{thebibliography}
\end{document}